 \documentclass[]{jfm}

\usepackage{graphicx}
\usepackage{amsmath}%
\usepackage{subfig}
\usepackage{psfrag}
\usepackage[usenames, dvipsnames]{color}
\usepackage{float}
\usepackage{natbib}
\usepackage{pifont}

\ifCUPmtlplainloaded \else
  \checkfont{eurm10}
  \iffontfound
    \IfFileExists{upmath.sty}
      {\typeout{^^JFound AMS Euler Roman fonts on the system,
                   using the 'upmath' package.^^J}%
       \usepackage{upmath}}
      {\typeout{^^JFound AMS Euler Roman fonts on the system, but you
                   dont seem to have the}%
       \typeout{'upmath' package installed. JFM.cls can take advantage
                 of these fonts,^^Jif you use 'upmath' package.^^J}%
      }
  \else
  \fi
\fi

\ifCUPmtlplainloaded \else
  \checkfont{msam10}
  \iffontfound
    \IfFileExists{amssymb.sty}
      {\typeout{^^JFound AMS Symbol fonts on the system, using the
                'amssymb' package.^^J}%
       \usepackage{amssymb}%
       \let\le=\leqslant  
       \let\ge=\geqslant  
      }{}
  \fi
\fi

\ifCUPmtlplainloaded \else
  \IfFileExists{amsbsy.sty}
    {\typeout{^^JFound the 'amsbsy' package on the system, using it.^^J}%
     \usepackage{amsbsy}}
    {}
\fi

\newcommand\Rey{\mbox{\textit{Re}}}  

\newsavebox{\astrutbox}
\sbox{\astrutbox}{\rule[-5pt]{0pt}{20pt}}

\newcommand{\diff}{\mathrm{d}}

 \title[]{Turbulence and secondary motions in square duct flow}

\author[S. Pirozzoli, D. Modesti, P. Orlandi, F. Grasso]{SERGIO PIROZZOLI$^1$, DAVIDE MODESTI$^2$, \\ PAOLO ORLANDI$^1$ AND FRANCESCO GRASSO$^{2}$} 

\affiliation{$^1$Dipartimento di Ingegneria Meccanica e Aerospaziale,
Sapienza Universit\`a di Roma \\ Via Eudossiana 18, 00184 Roma,
Italy \\ $^2$ Cnam-Laboratoire DynFluid, 151 Boulevard de L'Hopital, 75013 Paris\\[\affilskip]}

\pubyear{1996}
\volume{538}
\pagerange{119--126}
\date{\today}

\begin{document}

\maketitle

\begin{abstract}
We study turbulent flows in pressure-driven ducts with square cross-section 
through direct numerical simulation in a wide enough range of Reynolds number
to reach flow conditions which are representative of fully developed turbulence.
Numerical simulations are carried out over extremely long integration times
to get adequate convergence of the flow statistics, and specifically 
high-fidelity representation of the secondary motions which arise.
The intensity of the latter is found to be in the order of 1-2\% of
the bulk velocity, and unaffected by Reynolds number variations.
The smallness of the mean convection terms in the streamwise 
vorticity equation points to a simple characterization of the secondary flows,
which in the asymptotic high-$\Rey$ regime are found to be approximated
with good accuracy by eigenfunctions of the Laplace operator.
Despite their effect of redistributing the wall shear stress along the
duct perimeter, we find that secondary motions do not have large influence on the
mean velocity field, which can be characterized with good accuracy as that
resulting from the concurrent effect of four independent flat walls,
each controlling a quarter of the flow domain.
As a consequence, we find that parametrizations based on the hydraulic diameter 
concept, and modifications thereof, are successful in predicting the 
duct friction coefficient.
\end{abstract}

\section{Introduction}\label{sec:intro}

Internal flows within straight ducts having non-circular cross-section 
are common in many engineering applications,
such as water draining or ventilation systems, nuclear reactors,
heat exchangers and turbomachinery.
Within this class of flows, square ducts have attracted 
most of the interest, and these are the subject of this study.
Pioneering studies of flows in duct with complex cross-sections are due to
\citet{prandtl_27,nikuradse_30},
who first highlighted the presence of
secondary motions in the cross-stream plane, hence explaining the typical bending of the
mean streamwise velocity iso-lines towards the duct corners. 
Although the intensity of the secondary flow is a small 
fraction of the main stream (typically, a few percent), it may have some important practical impact, 
having the general role of redistributing friction and 
heat flux along the duct perimeter~\citep{leutheusser_63}.
In open-channel flows, secondary motions move fluid with relatively low
streamwise momentum towards the centre portion of the channel, hence causing a
depression of the velocity maximum below the free surface~\citep{nezu_05}. 
Furthermore, secondary motions play a major role in processes of sediment 
transport and river bed erosion~\citep{adrian_12}.
Hence, it is clear that robust physical insight and accurate prediction of flows
inside channels with complex shape is of utmost interest in the engineering practice.

Secondary flows are obviously associated with the presence of 
non-zero streamwise vorticity~\citep{bradshaw_87}
which in turn may arise because of skewing of existing spanwise vorticity
as in the case of duct bends (hence giving rise to Prandtl's secondary motions of the first kind),
and/or because of Reynolds stress gradients in the presence of non-circular 
cross-section (secondary motions of the second kind), which is the case of the present paper.
Secondary motions in square ducts are known to come in the form of eight counter-rotating eddies
bringing high-momentum fluid from the duct core towards the corners, and to have
a typical intensity of about $1\%$ the duct bulk velocity~\citep{prandtl_27}.
Modeling secondary flows is a challenging task for turbulence models,
as it is well known that classical models based on the isotropic eddy viscosity
ansatz cannot generate self-sustained secondary motions in straight ducts~\citep{speziale_82,mani_13}.
Quantitative measurements of secondary motions in ducts date back to the studies of
\citet{hoagland_60,brundrett_64,gessner_65}, who first attempted to shed light on the mechanisms
of vorticity generation and on the effect of Reynolds number variation.
Whereas those studies agree that secondary motions are generated from
gradients of the Reynolds stresses, \citet{brundrett_64} reported that convection of
mean streamwise vorticity provides an
important contribution to the overall vorticity balance, whereas~\citet{gessner_65}
reported convection to be at least one order of magnitude
less than the other balance terms in the mean vorticity equation.
\citet{brundrett_64} further argued that secondary eddies
should not be affected by the Reynolds number, whereas~\citet{gessner_65} found that
their intensity as a fraction of the bulk duct velocity decreases with increasing Reynolds number.
Consistent with the latter statement, \citet{launder_72} argued that the typical velocity 
scale of the secondary motions is the friction velocity, rather than the bulk velocity.
In a paper devoted to developing RANS closures for turbulent flows in ducts with
complex cross-section, \citet{demuren_84} pointed out that the contrasting conclusions of 
experimental studies regarding the Reynolds number dependence of secondary 
flows are likely due to incomplete flow development and/or inaccuracy of measurements.
From scrutiny of previous experimental data, those authors argued that
the longitudinal vorticity equations is controlled by balance between the 
Reynolds stress gradients and convection from the secondary motions,
and that the terms involving the difference of the normal stresses 
and of the secondary shear stress are comparable in magnitude.

Given the rather inconclusive outcome of experimental studies, 
it is clear that direct numerical simulation (DNS) 
may be a valuable tool to shed light on the nature of the secondary motions, as it
allows to accurately evaluate all the quantities which may be responsible for their
occurrence and sustainment, especially giving access to the near-wall region,
frequently disregarded in experiments owing to insufficient spatial resolution.
The first DNS of incompressible square duct flow was carried out by~\citet{gavrilakis_92}, at
bulk Reynolds number $\Rey_b=2hu_b/\nu=4410$ (where $h$ is the duct half side, $u_b$ is the 
bulk velocity in the duct, and $\nu$ is the fluid kinematic viscosity), which
corresponds to a friction Reynolds number $\Rey^*_{\tau}=hu^*_{\tau}/\nu=150$, 
where $u^*_\tau=(\tau^*_w/\rho)^{1/2}$ is the mean friction velocity, 
and $\tau^*_w$ is the mean wall shear stress.
Irrespective of the presence of secondary motions, the mean flow along the wall 
bisectors was found to be similar to the case of a plane channel. 
Contrary to the claims of \citet{demuren_84}, analysis of the mean 
streamwise vorticity equation showed that the gradients of the Reynolds stresses are approximately
balanced by viscous diffusion, whereas convection is less important.
This result was also confirmed by DNS at $\Rey^*_{\tau} \approx 300$ by \citet{huser_93}, 
although carried out with marginal spatial resolution.
\citet{uhlmann_07,pinelli_10} first attempted to numerically span
a range of (low) Reynolds numbers from 
$\Rey_{\tau}^* \approx 80$ to
$\Rey_{\tau}^* \approx 225$, also
noticing that much longer time integration intervals are necessary to
achieve convergence of the flow statistics than done in earlier studies.
Their study especially focused on establishing the spatial association
between streamwise vorticity and streamfunction
of the cross flow. Those authors noticed that as the Reynolds number increases
the peak locations of $\omega_x$ and $\psi$ start to be segregated, the former remaining
approximately constant in wall units, and the latter in outer units.
\citet{vinuesa_14} carried out DNS of duct flow in square and rectangular channels 
at $\Rey_{\tau}^*=180-300$, with special attention to establishing the effect of the channel aspect ratio.
Inner scaling with the local wall friction was found to yield greater
universality of the flow statistics as compared to scaling based on the mean wall friction.
\citet{zhang_15} carried out DNS of square duct flow up to $\Rey_{\tau}^*=600$, 
the highest reached so far.
They found that below $\Rey_{\tau}^*=300$, low-Reynolds number effects are dominant,
and observed a continuous trend in the position of the vortex centers, which move
towards the wall bisectors as the Reynolds number increases.
Although this was not commented in depth, their data seem to indicate growth of the intensity
of the secondary motions with the Reynolds number, when expressed in bulk units.
\citet{marin_16} carried out DNS of hexagonal duct flow up to $\Rey_{\tau}^*\approx400$
and compared the secondary motions with those observed in square duct flow, finding a similar trend
with the Reynolds number as observed by \citet{pinelli_10}, namely
that the streamfunction scales in outer units, whereas the streamwise vorticity scales in inner units.

In summary, although the flow in square ducts has been frequently studied both
through experiments and DNS, and it is used as a prototype for the study of flows 
in ducts with complex cross-section, it appears that several fundamental questions have
not been satisfactorily answered. The present paper then aims 
at elucidating the following issues:
i) which is the correct velocity scale for the secondary motions?;
ii) to what extent does the universal inner-layer law apply to the mean streamwise velocity profile?;
iii) what is the nature of the secondary motions, and is it possible to provide an
approximate characterization in the high-Reynolds-number limit?;
iv) what is the effect of the secondary motions on the bulk flow features, namely the friction law?
In order to answer these questions a novel DNS database has been developed
covering Reynolds number up to $\Rey_{\tau}^* \approx 1000$, as described in the following section.

\section{The numerical database}\label{sec:numerics}

\begin{figure}
 \begin{center}
  (a)
  \includegraphics[width=4.0cm,clip]{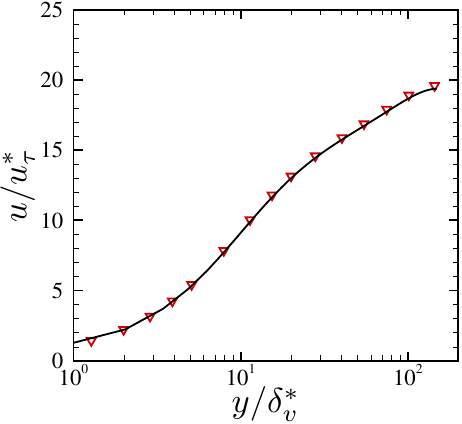}
  (b)
  \includegraphics[width=4.0cm,clip]{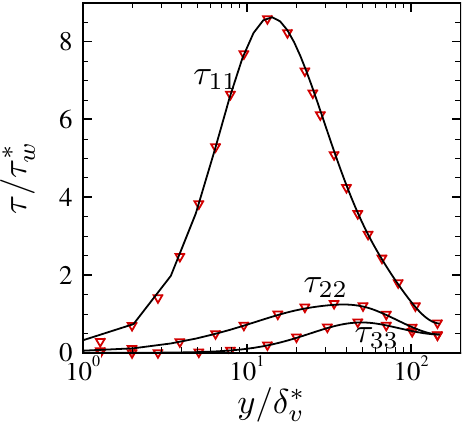}
\caption{Profiles of mean velocity (a) and turbulent normal stresses (b) 
         along wall bisector for flow case A (lines) compared with 
         DNS data of \citet{pinelli_10} (triangles). 
           }
  \label{fig:valid}
 \end{center}
\end{figure}

\begin{table}
\centering
\begin{tabular}{lccccccccccccc}
\hline
\hline
 Case & $\Rey_b$ & $\Rey_{\tau}^*$ & $C_f\times10^{-3}$ & $N_x$ & $N_y$ & $N_z$ & $\Delta x^*$ & $\Delta z^*$ & $\Delta y_w^*$ &${\Delta t}_{av}^* u^*_\tau/h$ & $\epsilon\times10^{-4}$ & $\epsilon_{s}\times10^{-2}$ \\
\hline
 A &$4410$ & $151$  & $9.38$ & $512$  & $128$ & $128$ & $5.5$ & $3.1$  & $0.71$ & $2204$ & $9.71$ & $4.10$ \\
 B &$7000$&  $225$  & $8.34$ & $640$  & $144$ & $144$  & $6.7$ & $5.1$ & $0.82$ & $1605$ & $2.69$ & $3.30$ \\
 C &$17800$& $516$  & $6.76$ & $1024$ & $256$ & $256$ & $9.6$ & $6.7$  & $0.88$ & $1385$ & $3.31$ & $2.30$ \\
 D &$40000$& $1048$ & $5.53$ & $2048$ & $512$ & $512$ & $9.7$ & $6.7$  & $0.63$ & $493$  & $1.69$ & $8.16$ \\
\hline
\end{tabular}
\caption{Flow parameters for square duct DNS.
Box dimensions are $6\pi h \times 2h \times 2h$ for all flow cases. 
$\Rey_b = 2 h u_b / \nu$ is the bulk Reynolds number, and 
$\Rey_{\tau}^* = h u_{\tau}^* / \nu$ is the friction Reynolds number.
$\Delta x$ is the mesh spacing in the streamwise direction, and 
$\Delta z$, $\Delta y_w$ are the maximum and minimum mesh spacings 
in the cross-stream direction, all given in global wall units, $\delta_v^*=\nu/u_{\tau}^*$.
${\Delta t}_{av}^*$ is the effective averaging time interval, and 
the convergence indices $\epsilon$ and $\epsilon_{s}$ are defined in 
equations~\eqref{eq.pdf}-\eqref{eq:eps_sym}.}
\label{tab:test}
\end{table}

The numerical simulation of turbulent flow in square ducts in the incompressible limit 
is a much more challenging task for numerical algorithms than the canonical cases of plane channel and pipe flow.
The main reason resides in the availability of only one direction of space inhomogeneity,
which prevents the use of efficient inversion procedures for Poisson equations
based on double trigonometric expansions~\citep{kim_85,orlandi_12}.
Although this difficulty can be circumvented through the use of two-dimensional Poisson 
solvers based on cyclic reduction~\citep{gavrilakis_92}, similar efficiency can also 
be achieved with compressible algorithms made to operate at low Mach number~\citep{modesti_16a}, 
thus avoiding the solution of a discrete Poisson equation, however at the expense of stiffness in the
allowed time step for explicit integration of acoustic waves.
In the present work we use a fourth-order co-located finite-difference solver, previously 
used for DNS of compressible turbulence, also in the low-Mach-number regime~\citep{pirozzoli_13,modesti_16}.
Here, the convective terms in the Navier-Stokes equations are preliminarily expanded to
quasi-skew-symmetric form, in such a way to discretely  
preserve total kinetic energy from convection~\citep{pirozzoli_10}. 
Semi-implicit time stepping is used for time advancement in order to relax the acoustic time step limitation, 
thus allowing efficient operation at low Mach number, also 
through the use of the entropy evolution equation rather than the total energy equation.
The streamwise momentum equation is forced in such a way as to maintain a constant mass flow rate
(the spatially uniform driving term is hereafter referred to as ${\Pi}$, see equation~\eqref{eq:mmb}), 
periodicity is exploited in the streamwise direction, and isothermal no-slip boundary 
conditions are used at the channel walls.

Let $h$ be the duct half-side, the DNS 
have been carried out for a duct with $[-h:h] \times [-h:h]$ cross section, 
and whose length is $6 \pi h$, the latter selected on the basis of
preliminary box size sensitivity studies as well as conclusions 
of previous authors~\citep{gavrilakis_92}.
Four DNS have been carried out at bulk Mach number $M_b=u_b/c_w=0.2$ (where $c_w$ is the speed of sound at
the wall temperature), and bulk Reynolds number $\Rey_b=4000-40000$ (see table~\ref{tab:test}), and hereafter labeled with letters from A to D.
The turbulence Mach number $M_t=u'/{c_w}$ nowhere exceeds $0.01$ for any of the simulations,
hence the present DNS may be regarded as representative of genuinely incompressible turbulence.
For the sake of later reference, we will use the $+$ superscript to denote quantities made nondimensional
with respect to the local wall friction, namely with $u_{\tau} = (\tau_w/\rho_w)^{1/2}$, 
$\delta_v=\nu/u_{\tau}$, and the $*$ superscript to denote quantities made nondimensional
with respect to the perimeter-averaged friction, $\tau^*_w = h \overline{\Pi}/2$,
$u^*_{\tau} = (\tau^*_w/\rho_w)^{1/2}$, $\delta_v^*=\nu/u_{\tau}^*$.

In order to validate the numerical approach, flow 
case A is made to match the conditions of previous DNS studies~\citep{gavrilakis_92,pinelli_10}. 
A comparison with the latter dataset is shown figure~\ref{fig:valid}, 
which supports excellent agreement of mean velocity and velocity fluctuation statistics.
Statistical convergence of DNS is a crucial issue,
and as pointed out by~\citet{oliver_14}, the statistical error
may be even dominant over the numerical error. 
In this respect, previous DNS of duct flow highlighted the need of extremely long
averaging time intervals to achieve statistical convergence, typically
several times longer than in plane channel flow. 
This issue is mainly associated with the weakness of secondary motions.
\citet{vinuesa_16} introduced a
convergence indicator for duct flow based on use of the mean momentum balance equation 
(see equation~\eqref{eq:mmb}) similar to what is usually done for plane channel flow, 
in which linearity of the total stress is taken as an index of statistical convergence. 
Specifically, those authors considered the root-mean-square 
of the residual of equation~\eqref{eq:mmb} expressed in wall units (say $R^*$)
and averaged over the wall bisector, 
\begin{equation}
 \epsilon = \left(\frac{1}{h}\int_0^h R^*(0,y)^2\mathrm{d}y\right)^{1/2}.
\label{eq.pdf}
\end{equation}
We have also considered an additional convergence indicator 
based on deviations of the computed statistical properties from the expected geometrical symmetry,
which we define as
\begin{equation}
 \epsilon_{s} = \frac{1}{u_b}\left[\frac{1}{4 h^2}\int_{-h}^h\int_{-h}^h (\overline{u}(y,z)-\overline{u}_{\mathrm{oct}}(y,z))^2\mathrm{d}y\mathrm{d}z\right]^{1/2},
\label{eq:eps_sym}
\end{equation}
where $\overline{u}_{\mathrm{oct}}$ denotes the mean velocity averaged over octants, 
as customarily reported by other authors.
Parameters relevant for the statistical convergence of DNS are given in table~\ref{tab:test}.
Following \citet{vinuesa_16}, in order to account for the effect of the streamwise length of the computational
domain on statistical convergence, we consider effective time averaging intervals,
$\Delta t_{av}^* = \Delta t_{av} L_x/(6 h)$, expressed in eddy-turnover times, $h/u_\tau^*$. 
For the sake of reference, previous studies used $\Delta t_{av}^*=52 h/u_\tau^*$~\citep{gavrilakis_92},
$\Delta t_{av}^*=114h/u_\tau^*$~\citep{pinelli_10}, $\Delta t_{av}^*=112 h/u_\tau^*$~\citep{vinuesa_14}.
The mean momentum balance convergence indicator is found to be less than $10^{-3}$ for 
all simulations, which is the convergence threshold suggested by \citet{vinuesa_14}.
The symmetry indicator further suggests that distortions with respect to a 
eight-fold symmetric state are no larger than a few percent.

\section{The secondary motions} \label{sec:second}

\begin{figure}
 \begin{center}
  (A)
  \includegraphics[width=5.0cm,clip]{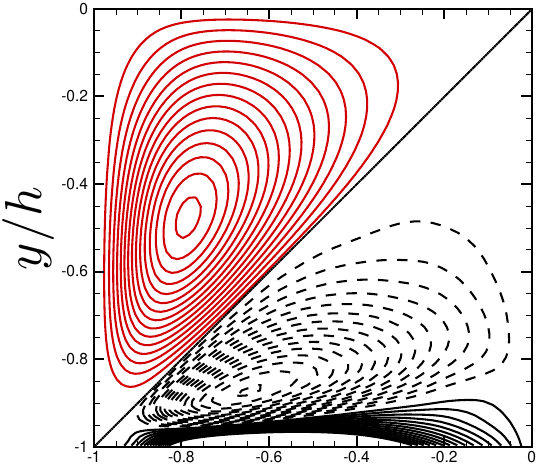}
  (B)
  \includegraphics[width=5.0cm,clip]{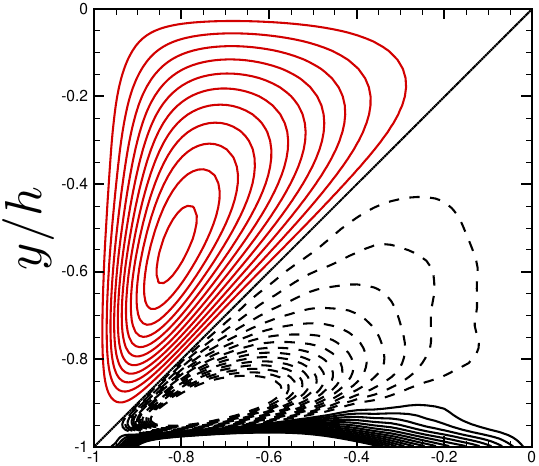}\\
  (C)
  \includegraphics[width=5.0cm,clip]{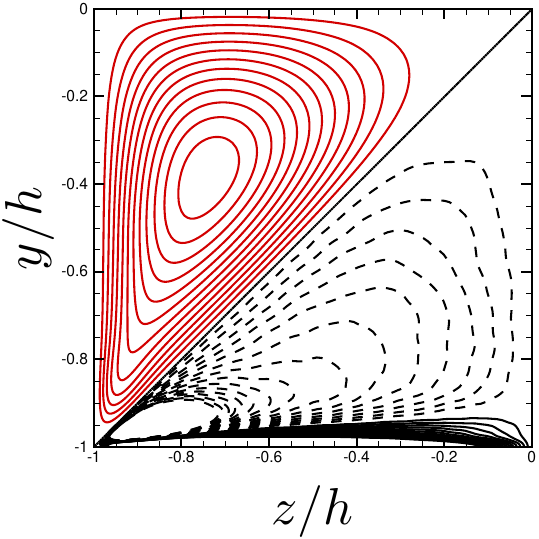}
  (D)
  \includegraphics[width=5.0cm,clip]{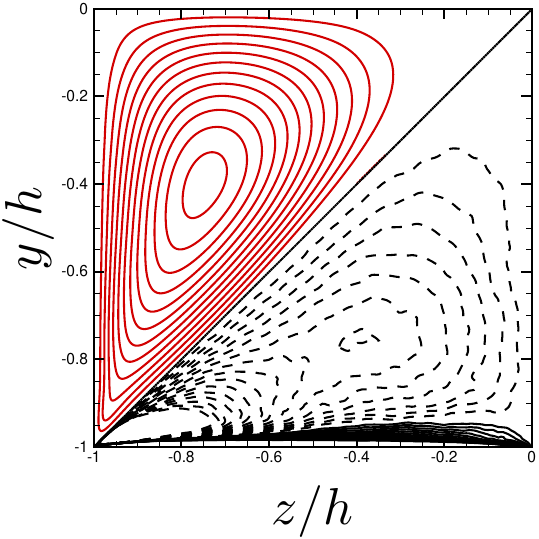}
  \caption{Vorticity contours (lower octant) and mean cross-flow streamlines (upper octant).
   Contours of $\overline{\omega}_x/(u_b/h)$ are shown from $-0.24$ to $0.24$, in intervals of $0.016$ (dashed lines denote negative values), and contours of $\psi/(u_b h)$ are shown from $0$ to $0.024$, in intervals of $0.0016$. Only one quadrant of the duct cross-section is shown. Refer to table~\ref{tab:test} for nomenclature of the flow cases.}
  \label{fig:psiomega}
 \end{center}
\end{figure}

A visual representation of the structure of the secondary motions is given in figure~\ref{fig:psiomega},
where we show the computed mean streamwise vorticity ($\overline{\omega}_x = \partial \overline{v} / \partial z - \partial \overline{w} / \partial y$) and the corresponding stream function, 
evaluated by solving 
\begin{equation}
\nabla^2 \psi = - \overline{\omega}_x . \label{eq:poisson}
\end{equation}
As well established~\citep{brundrett_64} the time-averaged secondary motions consist of eight eddies 
having triangular shape, with obvious symmetry properties, whose main effect is to 
redistribute momentum excess from the channel core towards the duct corners, where
momentum is less than the average because of the concurrent friction exerted by two walls.
The outer scaling ($u_b$, $h$) used in the figure allows to draw the first conclusion that the intensity of 
the eddies is approximately unaffected by Reynolds number variations. 
Although the streamline topology is not changing, it appears that the geometry of the eddies does in fact change in two ways.
First, the vortex centers exhibit a non-monotonic shift from the corners with the Reynolds number in a way to be discussed in detail later on, and second the streamlines penetrate deeper into the corners as $\Rey$ increases.
As also pointed out by \citet{pinelli_10}, the vorticity distribution is more complex than the streamline pattern. 
At low Reynolds number (A) a primary negative vorticity patch
is observed in the duct core, which is accompanied 
by a mirror layer of positive vorticity signed near the wall,
that forms as a result of the no-slip condition~\citep{orlandi_90}.
As the Reynolds number increases the main negative vorticity peak becomes stronger
in amplitude, and it becomes progressively confined toward the duct corner,
whereas the weaker core vorticity remains roughly of constant strength.
A secondary negative vorticity peak eventually emerges, which is visible
in (D) at about the same position as the 
centers of the cross-stream eddies.
Hence, it appears that the strongest vorticity 
(hereafter referred to as corner vorticity)
becomes progressively disconnected with the streamfunction distribution,
whereas the core vorticity, which scales well in outer units ($u_b/h$),
becomes closely associated with the streamfunction at high enough $\Rey$.

\begin{figure}
 \begin{center}
  (A)
  \includegraphics[width=4.0cm]{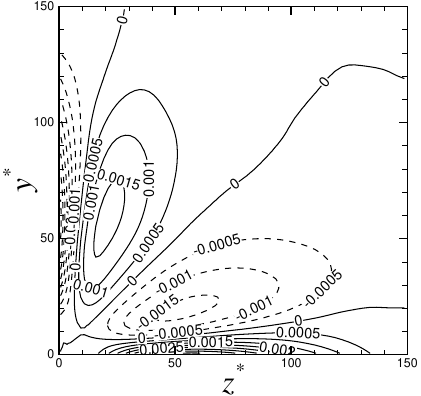}
  (B)
  \includegraphics[width=4.0cm]{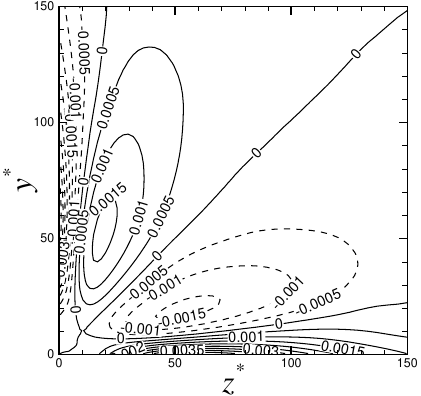} \\
  (C)
  \includegraphics[width=4.0cm]{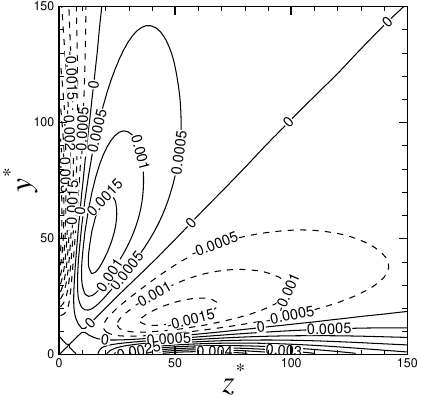}
  (D)
  \includegraphics[width=4.0cm]{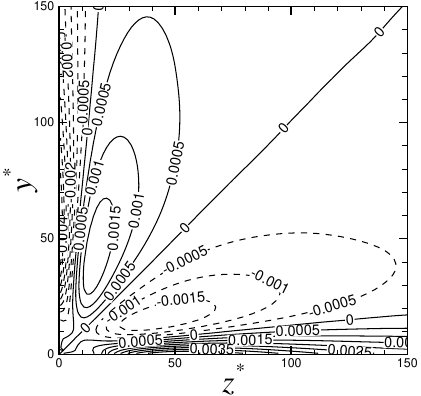}
  \caption{Streamwise vorticity contours near the bottom-left duct corner in inner coordinates, $z^*=(z+h)/\delta_v^*$, $y^*=(y+h)/\delta_v^*$. Contours are shown for $-0.004 \le \omega_x/(u_b/\delta_v^*) \le 0.004$, in intervals of $0.0005$ (dashed lines denote negative values).}
  \label{fig:omx_inn}
 \end{center}
\end{figure}


Deeper insight into the structure and scaling of the corner vorticity can be gained by
using an inner representation, as in figure~\ref{fig:omx_inn}. As noted in previous studies, the highest values of the
vorticity tend to be concentrated about the corner bisector and at the wall. These motions are empirically 
found to scale well in mixed units ($u_b, \delta_v^*$), thus they become progressively confined 
to the duct corners as $\Rey$ is increased, and their intensity increases when expressed in
outer units, as was clear in figure~\ref{fig:psiomega}. 
Despite its greater strength, this wall-confined 
vorticity does not contribute significantly to the secondary motions
at high $\Rey$. This is easily understood, since the total circulation associated with the 
core vorticity is expected to scale as $u_b h$, whereas the contribution of the
corner vorticity is expected to scale as $u_b \delta_v^*$, hence decreasing as 
$\Rey_{\tau}^*$ with respect to the former. This expectation is supported from the DNS data, which also shows that 
the ratio of the secondary (positive in figure~\ref{fig:psiomega}) circulation and the primary (negative) circulation is about $0.42$, regardless of the Reynolds numbers. 
A separate analysis (not reported here) further shows that the streamline pattern away from corners
is not significantly altered in shape and intensity if the corner vorticity is cut off
at the right-hand-side of equation~\eqref{eq:poisson}.

\begin{figure}
 \begin{center}
  (A)
  \includegraphics[width=4.0cm]{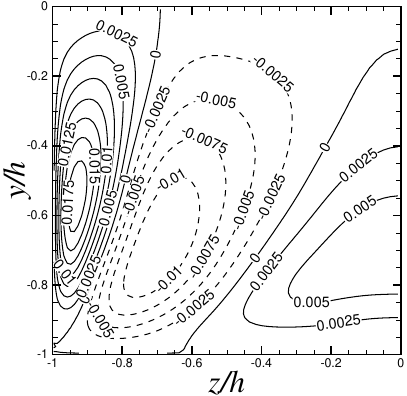}
  (B)
  \includegraphics[width=4.0cm]{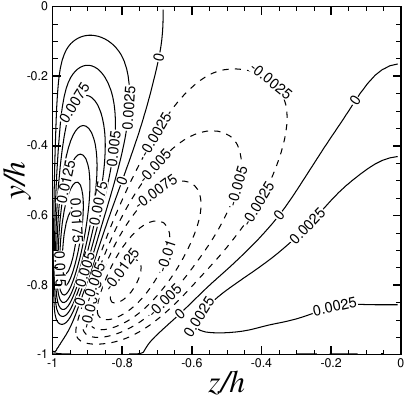} \\
  (C)
  \includegraphics[width=4.0cm]{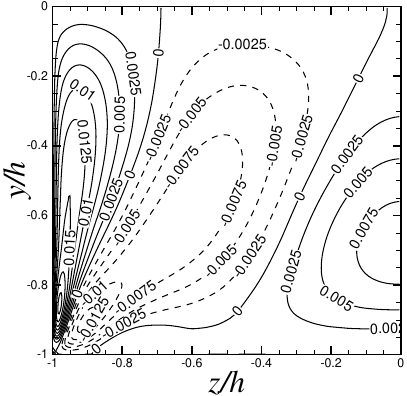}
  (D)
  \includegraphics[width=4.0cm]{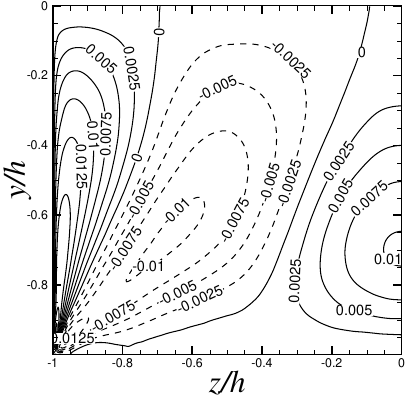}
  \caption{Contours of the mean cross-stream velocity component $\overline{v}$. Contour levels are shown for $-0.02 \le \overline{v}/u_b \le 0.02$, in intervals of $0.0025$ (dashed lines denote negative values).}
  \label{fig:v}
 \end{center}
\end{figure}

\begin{figure}
 \begin{center}
  (A)
  \includegraphics[width=4.0cm]{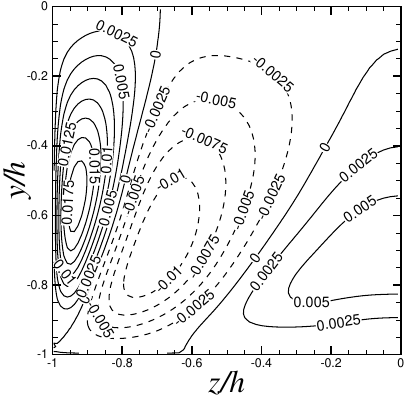}
  (B)
  \includegraphics[width=4.0cm]{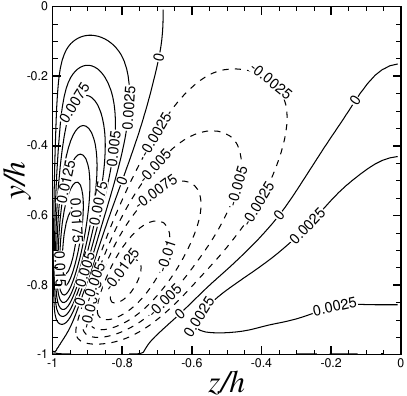} \\
  (C)
  \includegraphics[width=4.0cm]{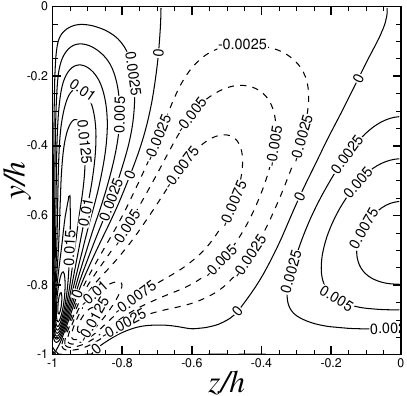}
  (D)
  \includegraphics[width=4.0cm]{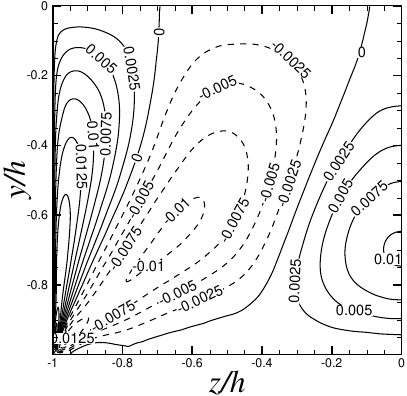}
  \caption{Contours of the mean cross-stream velocity component $\overline{v}$ near the duct corner in inner coordinates, $z^*=(z+h)/\delta_v^*$, $y^*=(y+h)/\delta_v^*$.
  Contour levels are shown for $-0.02 \le \overline{v}/u_b \le 0.02$, in intervals of $0.0025$ (dashed lines denote negative values).}
  \label{fig:v_inn}
 \end{center}
\end{figure}

Consistent with the previous observations, figure~\ref{fig:v} shows that the velocity associated 
with the secondary motions well scales with $u_b$, with maximum intensity of about $2\% u_b$. 
Relative maxima are attained for the vertical velocity
component along the corner bisector traced to the momentum inflow towards the duct corners, 
and parallel to the left sidewall and along the wall bisector as a consequence of the 
return motions to guarantee continuity. 
We may thus conclude that the correct scaling for the secondary flow is with outer units.
The observed scaling of the corner vorticity 
can then be explained as a viscous effect, associated with 
retardation of the secondary stream scaling on $O(u_b)$, and occurring
over a distance proportional to the viscous length scale $\delta_v^*$,
thus yielding the observed trend $\overline{\omega}_x \sim u_b u_{\tau}^*/\nu$.
As an alternative one might assume that the relevant
length scale in the duct corners is $\nu/u_b$, which would then lead to $\overline{\omega}_x \sim u_b^2/\nu$.
Based on the range of Reynolds number under scrutiny, the difference between the two trends 
is difficult to judge, however we are slightly in favour of the former option.
Figure~\ref{fig:v} also confirms that at sufficiently high $\Rey$ a new 
maximum of $v$ becomes isolated within the duct corners, which is associated 
with the corner vorticity. As clarified in figure~\ref{fig:v_inn}, 
the intensity of these motions also scales with $u_b$, and it is geometrically similar (on a smaller scale though)
to the core secondary flow.
An interesting scenario thus emerges, which includes two concurrent contributions to the
secondary motions: i) a core circulation, with typical velocity $u_b$ and typical
length scale $h$; ii) a corner circulation with typical velocity still $u_b$ and
typical length scale likely to be $\delta_v^*$. 
At low Reynolds number the two circulations basically coincide, whereas at higher 
Reynolds number the latter becomes progressively segregated to the duct
corners, and the core circulation emerges.




\begin{figure}
 \begin{center}
  (A)
  \includegraphics[width=4.0cm]{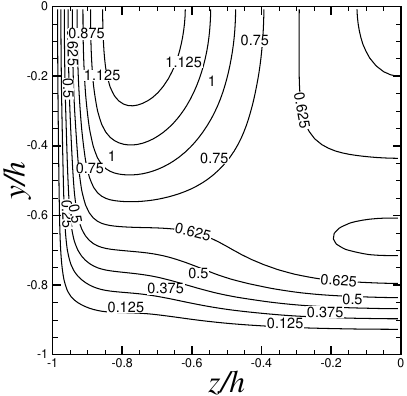}
  (B)
  \includegraphics[width=4.0cm]{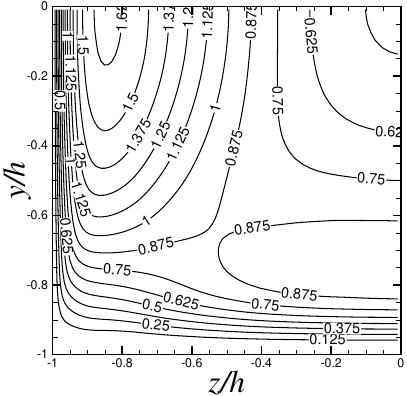} \\
  (C)
  \includegraphics[width=4.0cm]{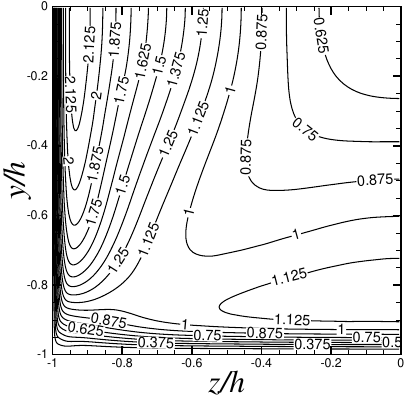}
  (D)
  \includegraphics[width=4.0cm]{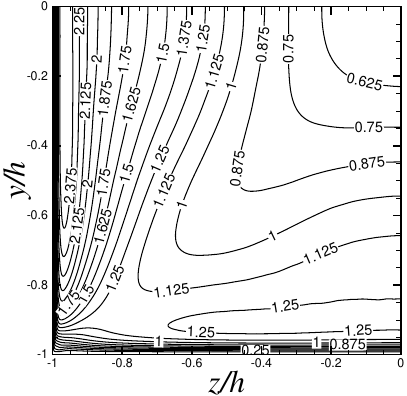}
  \caption{Contours of secondary turbulent normal stress, $0 \le \overline{v'^2}/{u^*_{\tau}}^2 \le 2$, in intervals of 0.125.}
  \label{fig:v1p_outer}
 \end{center}
\end{figure}




\begin{figure}
 \begin{center}
  (A)
  \includegraphics[width=4.0cm]{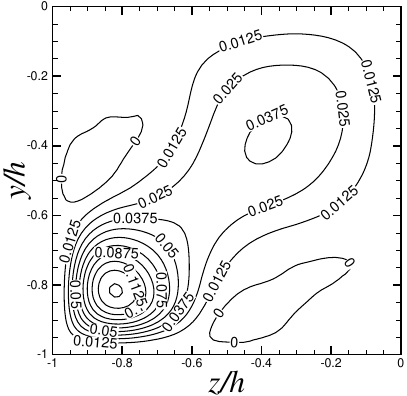}
  (B)
  \includegraphics[width=4.0cm]{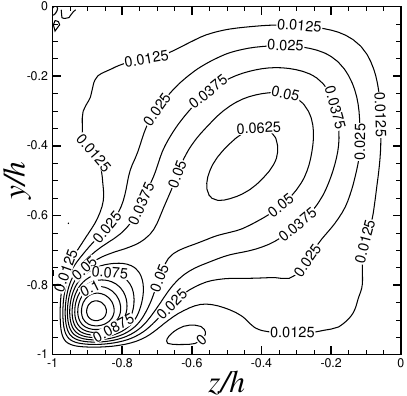} \\
  (C)
  \includegraphics[width=4.0cm]{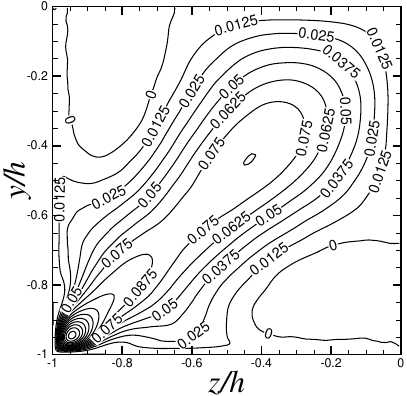}
  (D)
  \includegraphics[width=4.0cm]{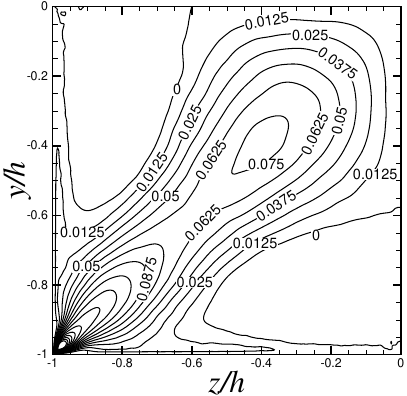}
  \caption{Contours of secondary turbulent shear stress, $0 \le \overline{v'w'}/{u^*_{\tau}}^2 \le 0.2$, in intervals of 0.0125.}
  \label{fig:v1w1p_outer}
 \end{center}
\end{figure}


Coming to the turbulent fluctuations, the distributions of the cross-stream turbulence 
intensity $\overline{v'^2}$ and of the cross-stream shear stress $\overline{v'w'}$ are
shown in figures~\ref{fig:v1p_outer}, \ref{fig:v1w1p_outer}. 
The turbulence intensity (figure~\ref{fig:v1p_outer}) 
is found to scale reasonably well in magnitude
with the mean friction velocity $u_{\tau}^*$, especially at the higher $\Rey$.
Near-wall peaks are observed
associated with local production from mean shear, which approach
the wall as the Reynolds number is increased.
It should be noted that the intensity is higher at the left wall,
with respect to which $v'$ plays the role of cross-stream velocity fluctuation, whereas
values are smaller at the bottom wall, where $v'$ has the role of a wall-normal fluctuation.
A ridge of $\overline{v'^2}$ is also observed along the corner bisector,
which penetrates deeper into the corner as the primary peaks recede toward the wall.
The secondary turbulent shear stress (figure~\ref{fig:v1w1p_outer}) 
has comparatively much smaller values (by at least 
an order of magnitude), and it mainly develops along the corner bisector.
Here again a two-scale organization is evident, whereby a relatively strong
peak becomes segregated to the duct corner, and a weaker peak 
associated with the main secondary circulation emerges.
In this case the scaling with $u_{\tau}^*$ is less clear than for $\overline{v'^2}$,
although the peak values seem to level off at the higher $\Rey$.


\begin{figure}
 \begin{center}
  (A)
  \includegraphics[width=3.6cm]{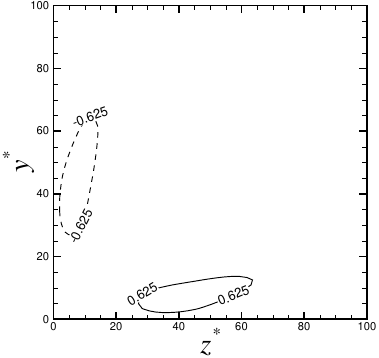}
  \includegraphics[width=3.6cm]{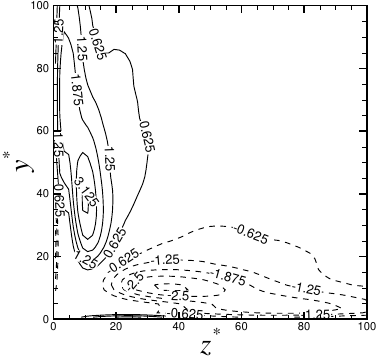}
  \includegraphics[width=3.6cm]{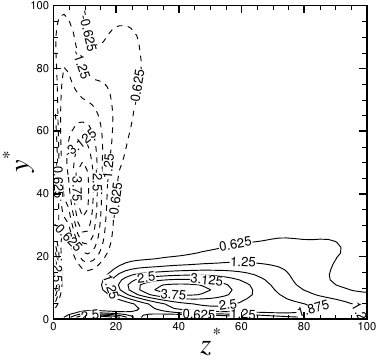} \\
  (B)
  \includegraphics[width=3.6cm]{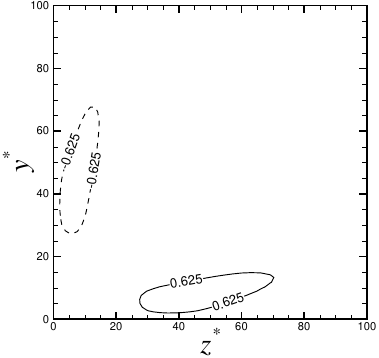}
  \includegraphics[width=3.6cm]{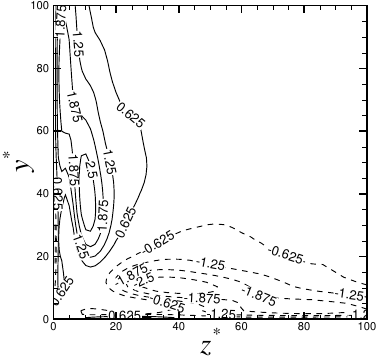}
  \includegraphics[width=3.6cm]{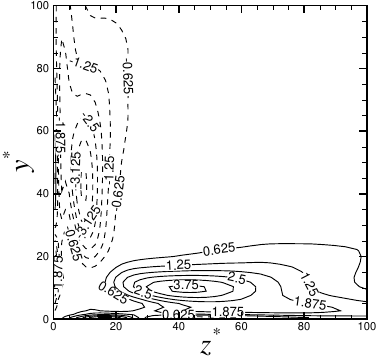} \\
  (C)
  \includegraphics[width=3.6cm]{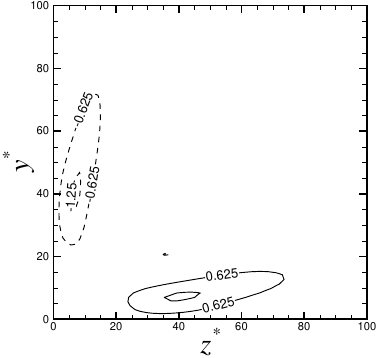}
  \includegraphics[width=3.6cm]{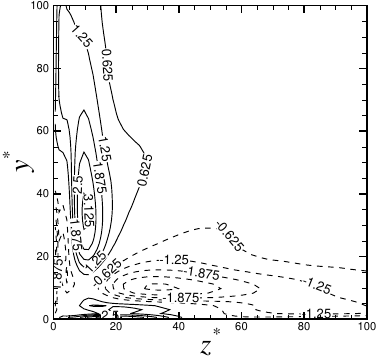}
  \includegraphics[width=3.6cm]{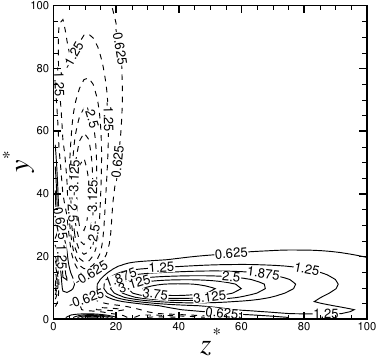} \\
  (D)
  \includegraphics[width=3.6cm]{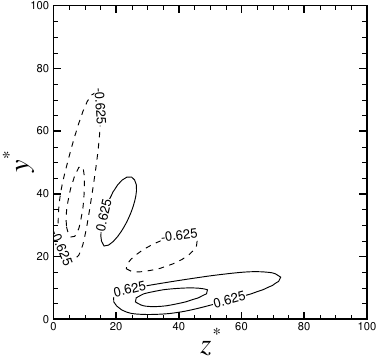}
  \includegraphics[width=3.6cm]{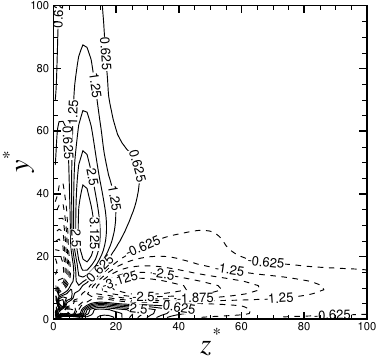}
  \includegraphics[width=3.6cm]{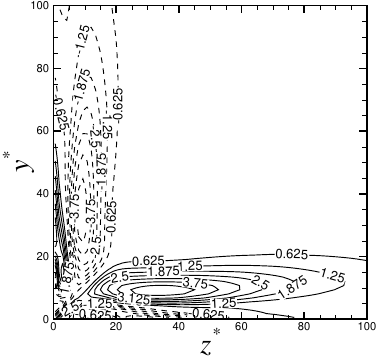} \\
  \caption{Terms in the mean streamwise vorticity budget equation (equation~\eqref{eq:vort}) near the duct corners: mean convection (left column), turbulence (middle column), viscous diffusion (right column). Contour levels are shown for $-5 \le (\cdot)/(u_b/\delta_v^*)^2 \cdot 10^6 \le 5$, in intervals of $0.625$ (dashed lines denote negative values).}
  \label{fig:omx_budg}
 \end{center}
\end{figure}

Deeper insight into the nature of the secondary motions can be gained by inspecting
the mean streamwise vorticity equation~\citep{einstein_58}
\begin{equation}
\overline{v}\frac{\partial\overline{\omega}_x}{\partial{y}} +
\overline{w}\frac{\partial\overline{\omega}_x}{\partial{z}}  =
\left(\frac{\partial^2}{\partial y^2}-\frac{\partial^2}{\partial z^2}\right)\left(-\overline{v'w'}\right) +
\frac{\partial^2}{\partial y\partial z}\left(\overline{v'^2}-\overline{w'^2} \right) +
\nu\left(\frac{\partial^2\overline{\omega}_x}{\partial y^2} + \frac{\partial^2\overline{\omega}_x}{\partial z^2} \right) .
\label{eq:vort}
\end{equation}
The various terms which appear in equation~\eqref{eq:vort} are associated with the effect
of mean cross-stream convection (left-hand side), secondary turbulent shear stress (first term at right hand side), normal stress anisotropy (second term), and viscous diffusion (third term).
Their distributions as determined from DNS are shown in figure~\ref{fig:omx_budg}, where 
the two turbulence terms are merged together, and all terms are normalized with respect to
$(u_b/\delta_v^*)^2$, which based on the previous discussion is the expected 
order of magnitude for the time derivative of $\overline{\omega}_x$. 
Also note that the distributions are only shown 
near the corners, where the various terms are sensibly different from zero.
The figure indeed supports the alleged scaling of the various terms across the given
Reynolds number range. Furthermore, it is indicative of
a leading balance between the turbulence terms 
(which are regarded as production terms in the common interpretation) and viscous diffusion,
whereas it appears that convection does not yield a significant contribution
to the streamwise vorticity dynamics. This scenario is consistent with low-$\Rey$ DNS studies~\citep{gavrilakis_92},
but it contradicts the typical conclusions of experimental studies,
which however are affected by significant uncertainties~\citep{demuren_84}.

\begin{figure}
 \begin{center}
  (a)
  \includegraphics[width=3.6cm]{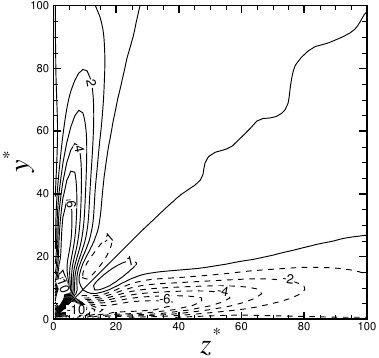}
  (b)
  \includegraphics[width=3.6cm]{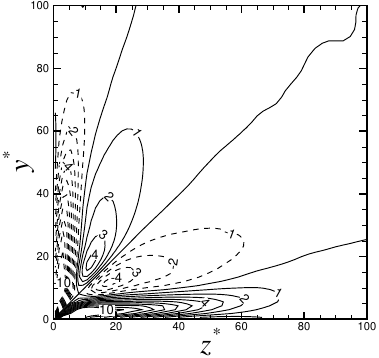}
  (c) 
  \includegraphics[width=3.6cm]{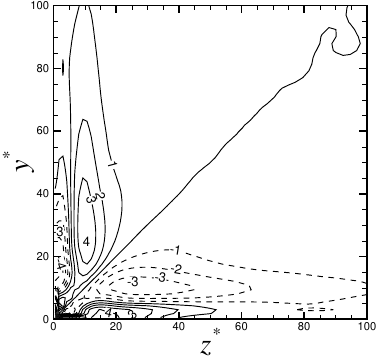}
  \caption{Turbulence terms in the mean streamwise vorticity budget equation (equation~\eqref{eq:vort}) for flow case D: anisotropy term (a) secondary shear stress term (b), and their sum (c). Contour levels are shown for $-10 \le (\cdot)/(u_b/\delta_v^*)^2 \cdot 10^6 \le 10$, in intervals of $0.5$ (dashed lines denote negative values).}
  \label{fig:omx_turb}
 \end{center}
\end{figure}

More detailed information on the individual contribution of the anisotropy and the
secondary shear stress terms is given in figure~\ref{fig:omx_turb}.
It is found that the two terms are overall of the same order of magnitude, although
the former is generally only considered in attempts of turbulence closure~\citep{demuren_84}.
Detailed inspection of the figure shows that their spatial organization is similar,
but they tend to be locally opposite in sign, with strong cancellation
near the walls, where both terms assume individually very large values. 
Comparing with their algebraic sum 
(repeated in panel (c) for convenience), it appears that non-zero values of 
turbulence production are primarily associated with large values of the secondary shear stress term.

%



\section{Effect on the mean streamwise velocity field} \label{sec:mean}

\begin{figure}
 \begin{center}
  (A)
  \includegraphics[width=5.0cm,clip]{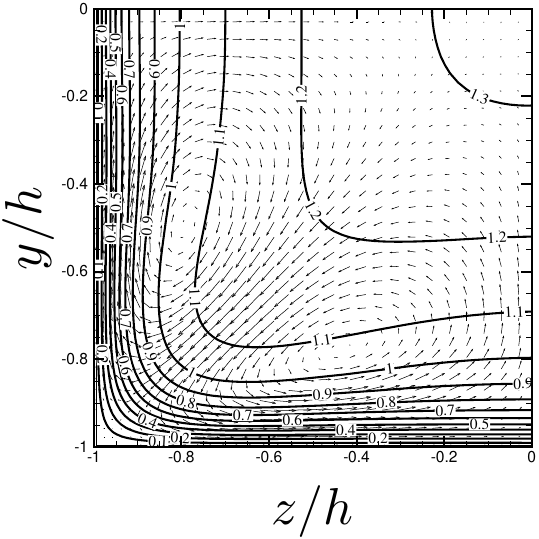}
  (B)
  \includegraphics[width=5.0cm,clip]{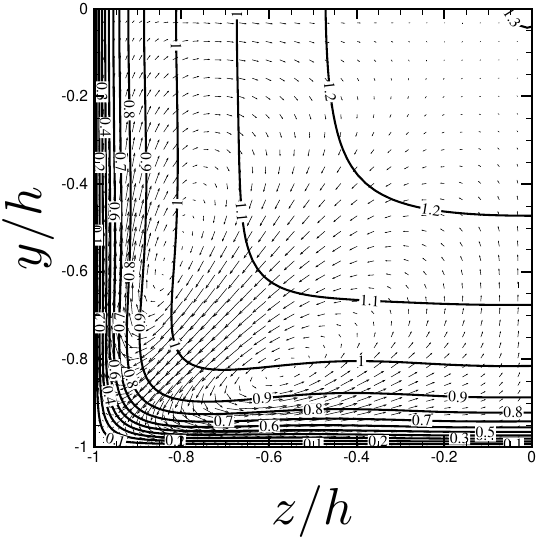}\\
  \vspace{1em}
  (C)
  \includegraphics[width=5.0cm,clip]{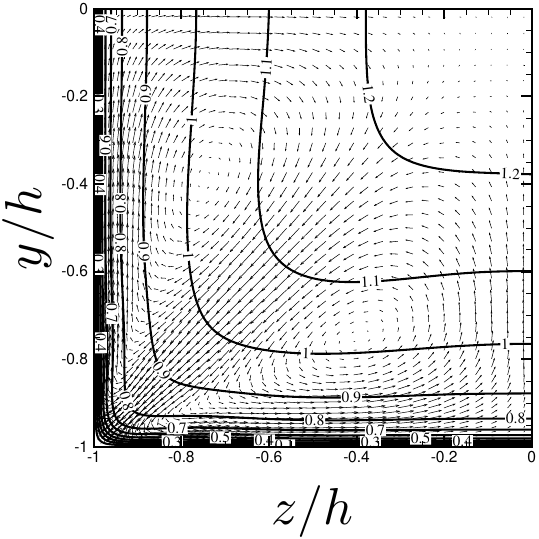}
  (D)
  \includegraphics[width=5.0cm,clip]{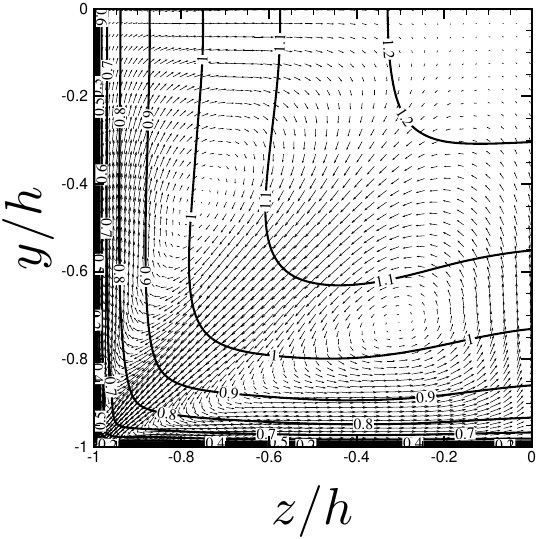}
  \vspace{1em}
   \caption{Mean streamwise velocity contours ($0 \le \overline{u}/u_b \le 1.3$) with superposed mean cross-stream vectors.}
  \label{fig:cross}
 \end{center}
\end{figure}

A representation of the organization of the flow field in cross-stream traverses
is provided in figure~\ref{fig:cross}, where we show mean velocity vectors superposed 
to iso-contour lines of the mean streamwise velocity. Consistent with previous studies,
we also observe the velocity contours to bulge toward the corners, under the action of the
main secondary currents heading wall-ward along the corner bisector.
On the other hand, the wall layer is found to thicken toward the wall bisector
under the action of the return currents heading away from the wall.
We find that the amount of distortion of the streamwise velocity iso-lines is non-monotonic
with the Reynolds number, with minimum distortion observed for flow case B.
This non-monotonic behavior is likely due to post-transitional effects, but
based on the previous observation that the strength of the secondary motions is 
proportional to $u_b$, we expect distortions of $\overline{u}$ to saturate at high enough Reynolds number.

\begin{figure}
 \begin{center}
  (a)
  \includegraphics[width=4.4cm,clip]{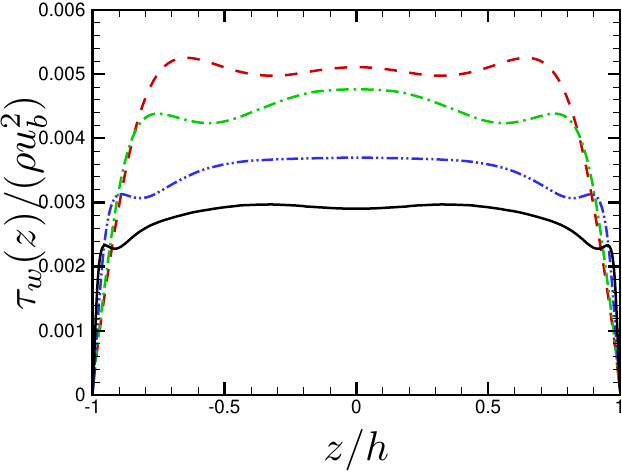}
  (b)
  \includegraphics[width=4.4cm,clip]{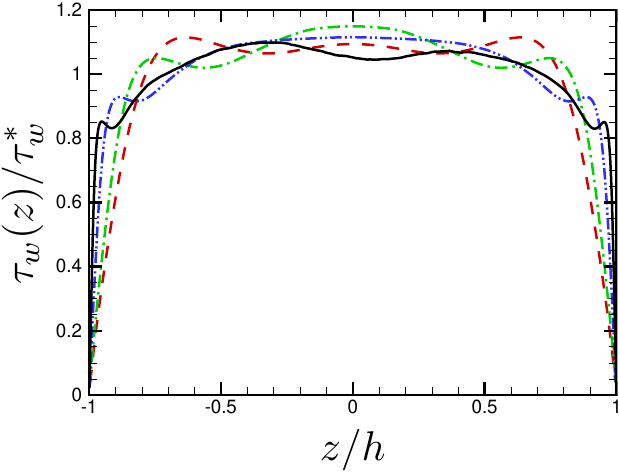}
  \caption{Local wall shear stress along the bottom wall, normalized with respect to 
           the dynamic pressure (a), and with respect to its mean value (b)
           for flow case A (dashed), B (dash-dot), C (dash-dot-dot), D (solid).}
  \label{fig:tauw}
 \end{center}
\end{figure}

The modulating effect of the secondary motions on the streamwise velocity field
is made more evident from inspection of the local wall friction distribution, see figure~\ref{fig:tauw}.
Rather than being uniform along the duct side, the local
friction must obviously drop to zero towards the corners. A non-monotonic trend is 
also evident here, with maximum values of the skin friction which are first attained 
midway between the corner and the wall bisector (flow case A), then at the wall bisector (flow cases B, C), 
and then again in between at higher Reynolds number. 
The behavior observed in flow case D is qualitatively very similar to what
found in experimental measurements~\citep{leutheusser_84}, generally carried out at
higher Reynolds number, and it is consistent with the previously
observed shape of the velocity iso-lines (see figure~\ref{fig:cross}).
In any case, disregarding the corner region, whose region of influence 
shrinks with $\Rey$, figure~\ref{fig:tauw} shows that deviations of the local shear stress from its 
mean value are no larger than $10\%$.

\begin{figure}
 \begin{center}
  (A)
  \includegraphics[width=5.0cm,clip]{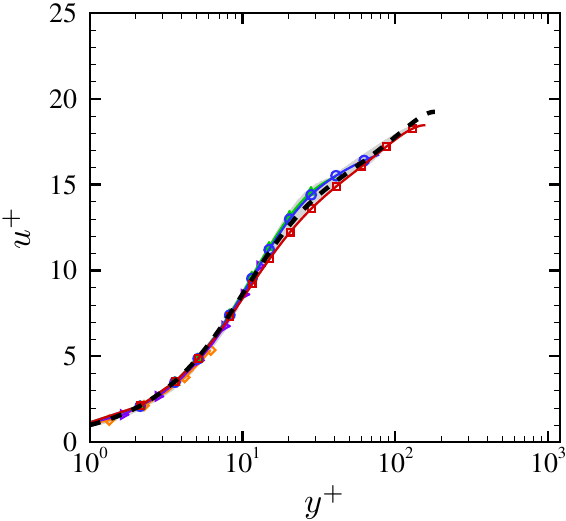}
  (B)
  \includegraphics[width=5.0cm,clip]{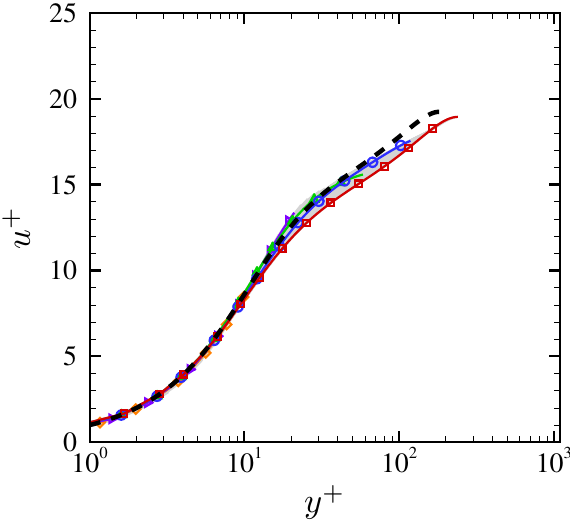} \\
  (C)
  \includegraphics[width=5.0cm,clip]{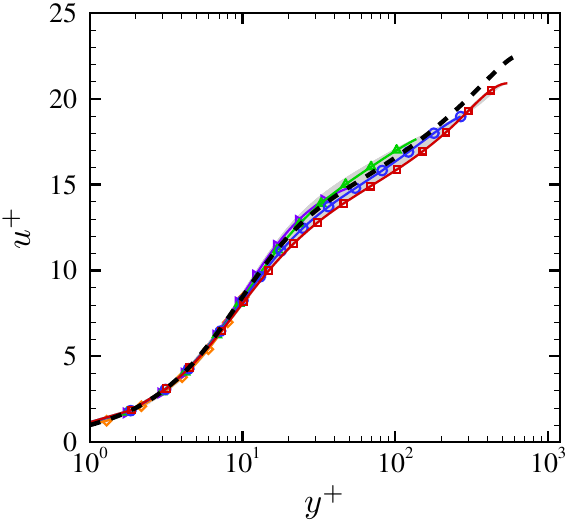}
  (D)
  \includegraphics[width=5.0cm,clip]{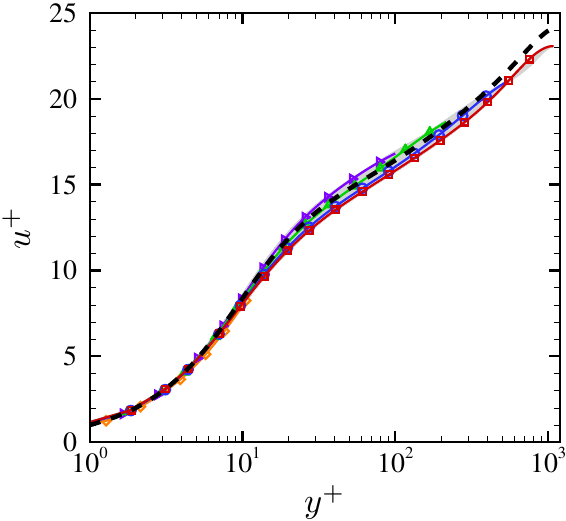}
  \caption{Mean streamwise velocity profiles along the $y$ direction (up to the corner bisector), given in local wall units at all $z$.
           Representative stations along the bottom wall are highlighted, namely 
           $z^*=15$ (diamonds), 
           $(z+h)/h=0.1$ (right triangles),
           $(z+h)/h=0.25$ (triangles),
           $(z+h)/h=0.5$ (circles), 
           $(z+h)/h=1$ (squares).
           The dashed lines denote mean profiles from DNS of pipe flow at
           $\Rey_{\tau}=181$ (A-B),  
           $\Rey_{\tau}=685$ (C),  
           $\Rey_{\tau}=1142$ (D), from~\citet{wu_08}.  
           }
  \label{fig:vel_prof}
 \end{center}
\end{figure}

\begin{figure}
 \begin{center}
  (A)
  \includegraphics[width=5.0cm,clip]{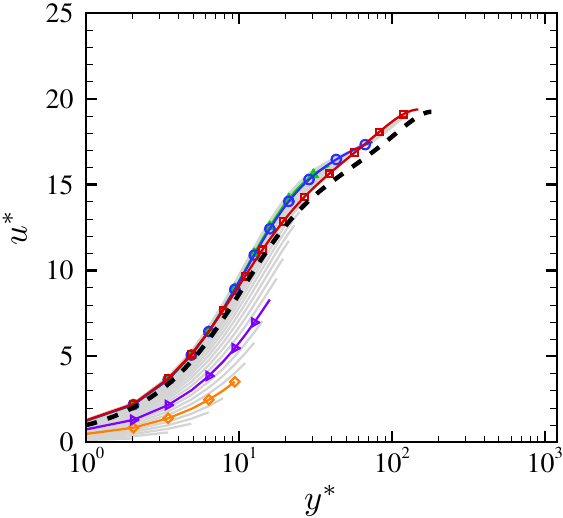}
  (B)
  \includegraphics[width=5.0cm,clip]{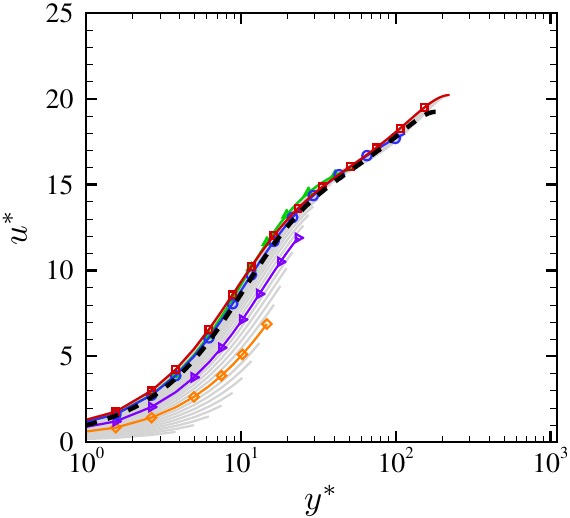} \\
  (C)
  \includegraphics[width=5.0cm,clip]{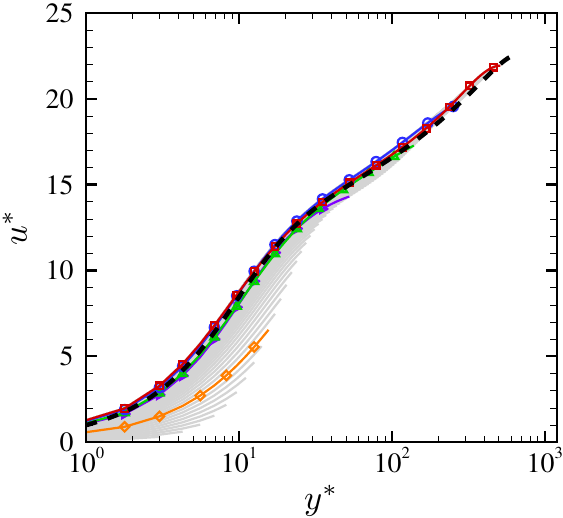}
  (D)
  \includegraphics[width=5.0cm,clip]{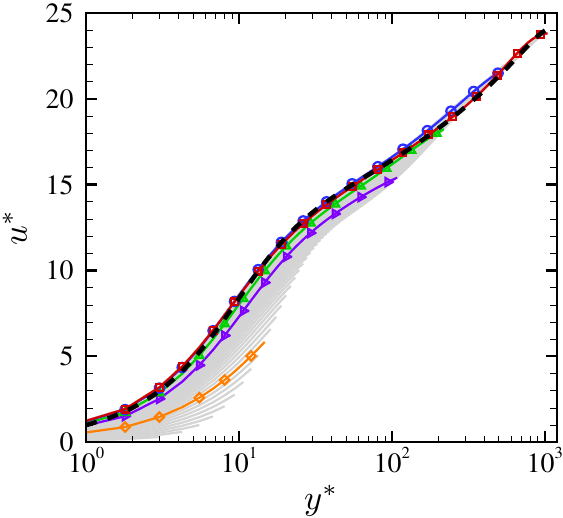}
  \caption{Mean streamwise velocity profiles along the $y$ direction (up to the corner bisector), given in global wall units at all $z$.
           Representative stations along the bottom wall are highlighted, namely 
           $z^*=15$ (diamonds), 
           $z/h=0.1$ (right triangles),
           $z/h=0.25$ (triangles),
           $z/h=0.5$ (circles), 
           $z/h=1$ (squares).
           The dashed lines denote mean profiles from DNS of pipe flow at
           $\Rey_{\tau}=181$ (A-B),  
           $\Rey_{\tau}=685$ (C),  
           $\Rey_{\tau}=1142$ (D), from~\citet{wu_08}.  
           }
  \label{fig:vel_prof_star}
 \end{center}
\end{figure}

\begin{figure}
 \begin{center}
  (A)
  \includegraphics[width=5.0cm,clip]{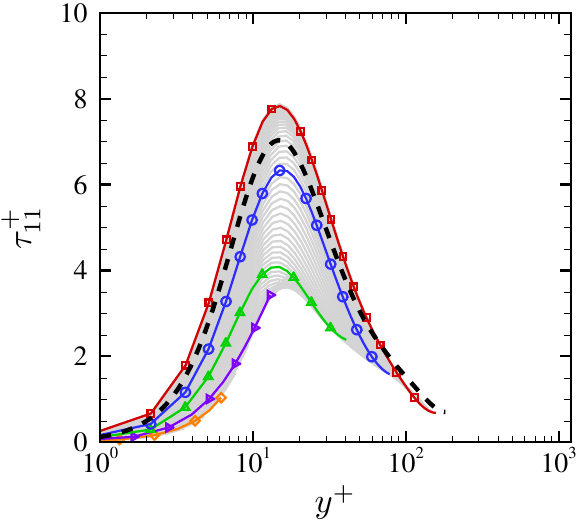}
  (B)
  \includegraphics[width=5.0cm,clip]{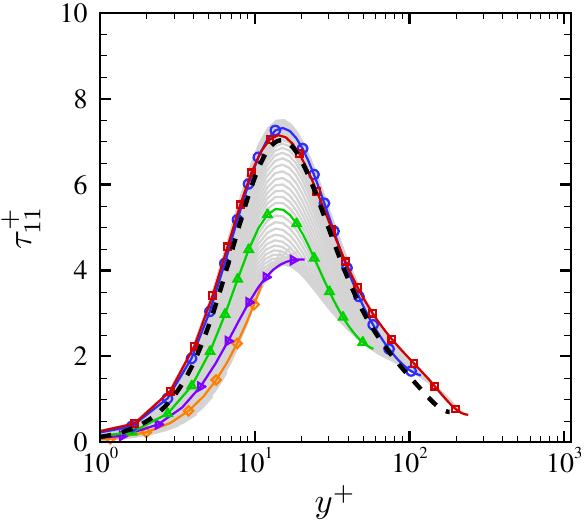} \\
  (C)
  \includegraphics[width=5.0cm,clip]{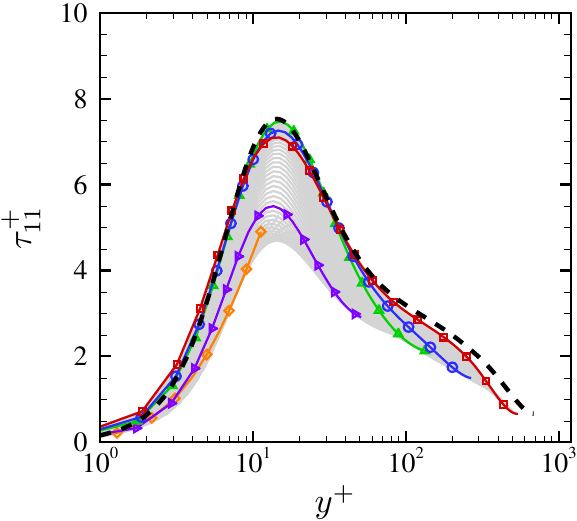}
  (D)
  \includegraphics[width=5.0cm,clip]{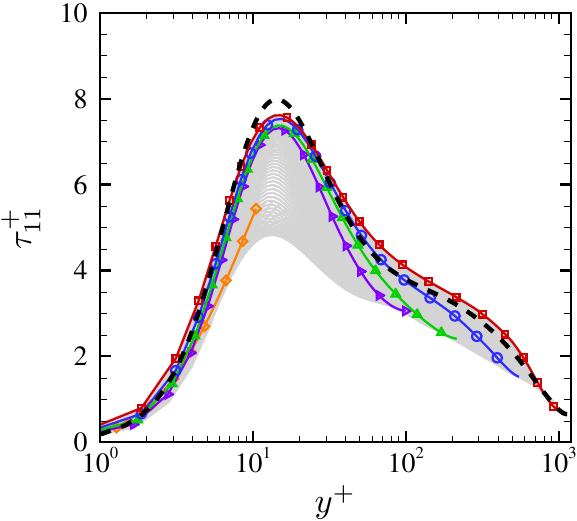}
  \caption{Mean streamwise turbulence intensity profiles along the $y$ direction (up to the corner bisector), given in global local units at all $z$.
           Representative stations along the bottom wall are highlighted, namely 
           $z^*=15$ (diamonds), 
           $z/h=0.1$ (right triangles),
           $z/h=0.25$ (triangles),
           $z/h=0.5$ (circles), 
           $z/h=1$ (squares).
           The dashed lines denote mean profiles from DNS of pipe flow at
           $\Rey_{\tau}=181$ (A-B),  
           $\Rey_{\tau}=685$ (C),  
           $\Rey_{\tau}=1142$ (D), from~\citet{wu_08}.  
           }
  \label{fig:tau11_prof}
 \end{center}
\end{figure}

Based on the previous observations, it is natural to study and compare the velocity statistics 
expressed in local wall units ($+$) and in global wall units ($*$).
Figure~\ref{fig:vel_prof} shows the mean velocity profiles as a function of the
wall-normal distance up to the corner bisector (where mean velocity has a maximum), 
in local wall units. For reference purposes, the mean velocity profiles from DNS of pipe 
flow at approximately matching $\Rey_{\tau}$ are also reported~\citep{wu_08}.
Excellent collapse of the locally scaled profiles is recovered near the wall,
also including the near-corner region.
The distributions become more widespread past $y^+ \approx 10$,
with maximum scatter of about $\pm 5\%$ in the ``logarithmic'' layer.
When the mean friction velocity is used for normalization
(see figure~\ref{fig:vel_prof_star}), 
scatter among the velocity profiles is observed 
near the wall as a result of the variation of the wall shear stress.
Perhaps unexpectedly, this normalization does yield better collapse 
of the various curves further away from the wall, and near coincidence
with the pipe velocity profiles, at least at high enough $\Rey$. 
This finding is probably related to the fact that mean momentum balance 
(as from equation~\eqref{eq:mmb}) is controlled by the imposed spatially uniform pressure gradient
rather than by the nonuniform wall shear. Hence
it is reasonable that, at least sufficiently away from walls,  
the flow should respond to the imposed pressure gradient rather than to the wall underneath.
Inspection of figure~\ref{fig:vel_prof_star} shows that transition from wall scaling
to pressure scaling occurs at a wall distance of about $0.2 h$, 
which is also the lower limit for the core region in canonical flows~\citep{pope_00}.
This interesting effect is impossible to observe in canonical wall-bounded flows,
in which the wall shear stress is spatially uniform.

The inner-scaled distributions of the streamwise Reynolds stress are shown 
in figure~\ref{fig:tau11_prof}. 
Along most of the wall, the behavior is qualitatively similar to canonical channel flow,
with a near-wall peak of $u'$ at $y^+ \approx 12$. 
The scatter among the various sections appears to be generally much larger 
than for the mean velocity field, although it probably becomes confined
to the corner vicinity at high $\Rey$. 
An interesting feature is the behavior of the peak streamwise normal stress
with the Reynolds number, which is larger at low $\Rey$ and then levels off
at increasing $\Rey$. This is apparently contradicting the 
increasing logarithmic trend of the wall-parallel velocity variances 
observed in canonical flows~\citep{bernardini_14}, 
hence it would be interesting to carry out DNS at yet higher Reynolds number 
to verify whether the present observations are related to 
viscous effects, or whether mechanisms of inner-outer layer interactions
in square ducts are different.

\begin{figure}
 \begin{center}
  (A)
  \includegraphics[width=3.6cm]{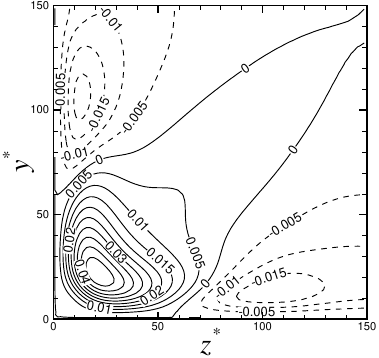}
  \includegraphics[width=3.6cm]{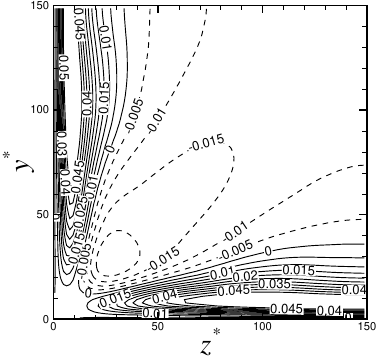}
  \includegraphics[width=3.6cm]{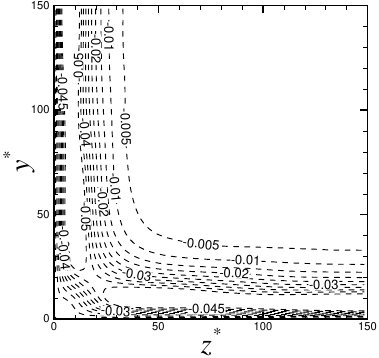} \\
  (B)
  \includegraphics[width=3.6cm]{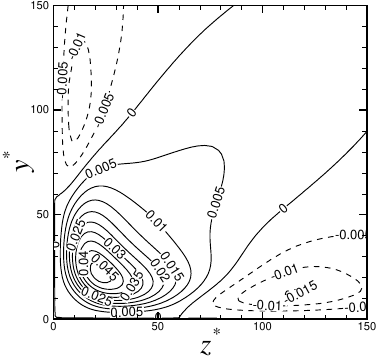}
  \includegraphics[width=3.6cm]{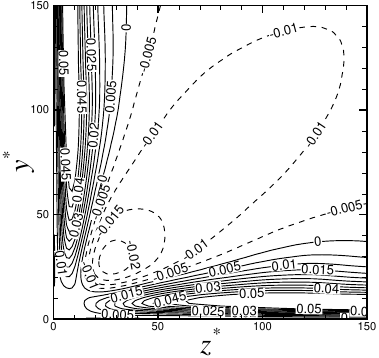}
  \includegraphics[width=3.6cm]{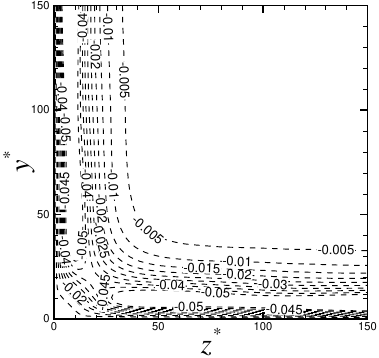} \\
  (C)
  \includegraphics[width=3.6cm]{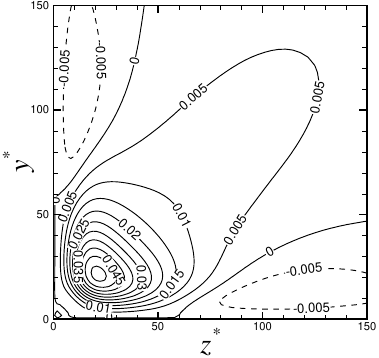}
  \includegraphics[width=3.6cm]{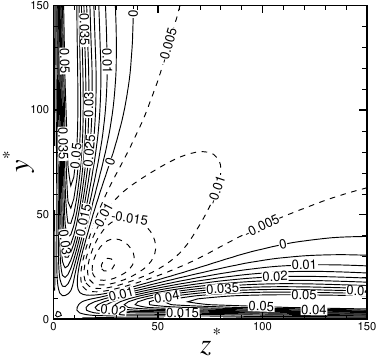}
  \includegraphics[width=3.6cm]{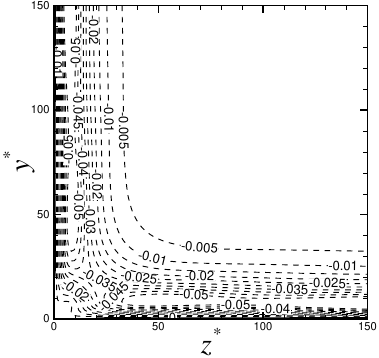} \\
  (D)
  \includegraphics[width=3.6cm]{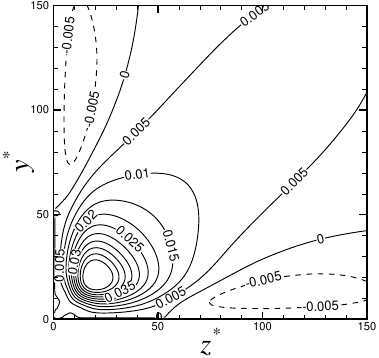}
  \includegraphics[width=3.6cm]{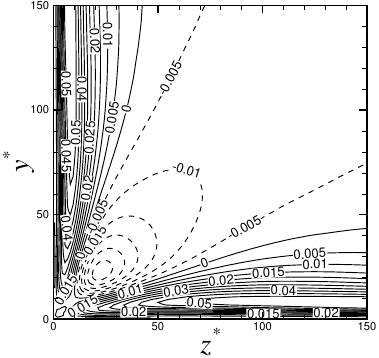}
  \includegraphics[width=3.6cm]{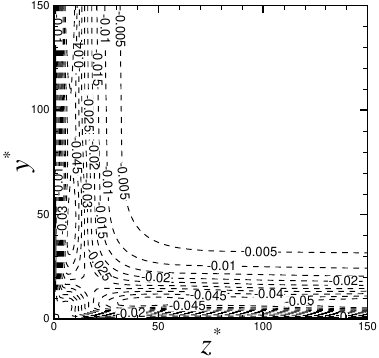} \\
  \caption{Terms in the mean streamwise momentum budget equation~\eqref{eq:mmb}: convection (left column),
           turbulence (middle column), viscous diffusion (right column). 
           Contour levels are shown for $-0.05 \le (\cdot)/({u_{\tau}^*}^2/\delta_v^*)^2 \le 0.05$, in intervals of $0.005$ (dashed lines denote negative values).}
  \label{fig:mom_budg}
 \end{center}
\end{figure}

The terms in the budget of the mean streamwise momentum equation,
\begin{equation}
\overline{v}\frac{\partial\overline{u}}{\partial y} + 
\overline{w}\frac{\partial\overline{u}}{\partial z} =
- \frac{\partial\overline{u'v'}}{\partial y} 
- \frac{\partial\overline{u'w'}}{\partial z}
+ \nu\left(\frac{\partial^2\overline{u}}{\partial y^2}  
+ \frac{\partial^2\overline{u}}{\partial z^2}\right) 
- \overline{\Pi} .
\label{eq:mmb}
\end{equation}
are shown in figure~\ref{fig:mom_budg}. Here, the left-hand-side represents the contribution of
convection from the secondary flows, and the terms at the right-hand-side represent the effects
of Reynolds stress gradients and viscous diffusion, and the driving pressure gradient.
Under global wall scaling (namely, each term is normalized with respect to ${u_{\tau}^*}^2/\delta_v^*$),
the magnitude of the various terms in the budget is approximately unaffected by $\Rey$ variation. 
Away from corners, the leading balance is between the wall-normal 
turbulent shear stress gradients and viscous diffusion as in canonical flows, whereas
mean cross-stream convection is at most one order of magnitude smaller.
Convection appears to play a role near corners, where it is responsible for positive
momentum transfer, and where it is balanced partly by diffusion and partly by locally negative
turbulent stress gradient. Hence, this scenario corroborates the previous findings about 
the mean velocity profiles (figures \ref{fig:vel_prof}-\ref{fig:vel_prof_star}). 
Since convection has a minor effect, a logarithmic layer is 
expected to arise at sufficiently high $\Rey$, excepted in the close vicinity of corners.

\begin{figure}
 \begin{center}
  \includegraphics[width=7.0cm]{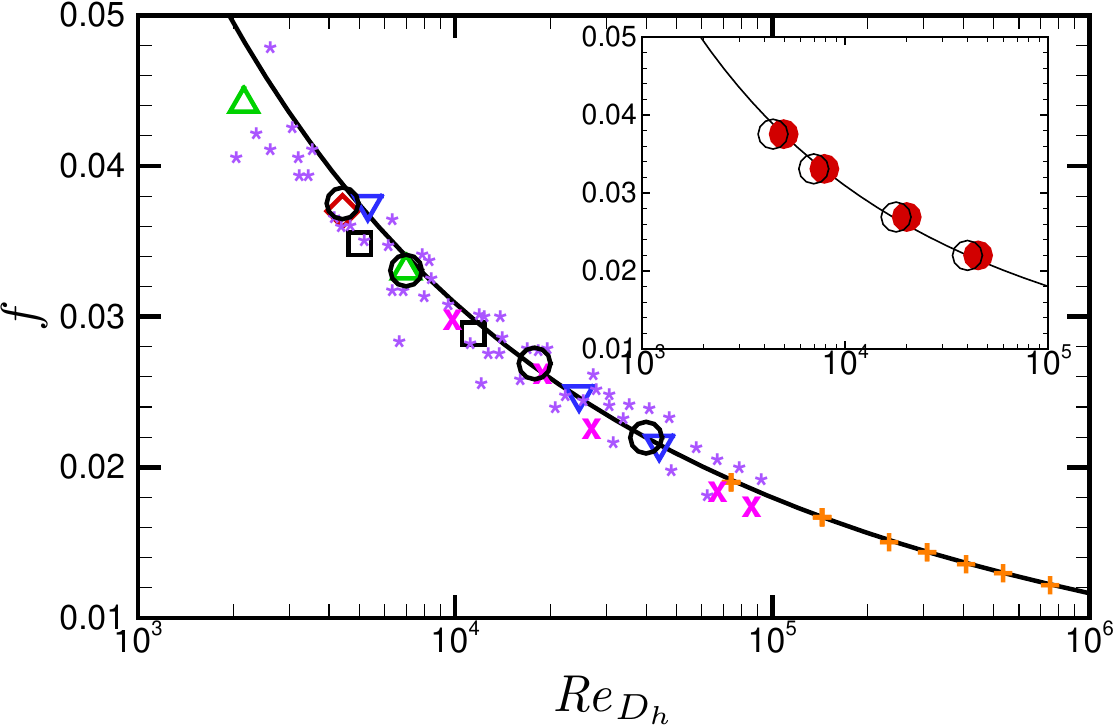}
  \vskip1.em
  \caption{Friction factor ($f = 4 C_f$) as a function of bulk Reynolds number based on the hydraulic diameter ($\Rey_{D_h}=u_b D_h/\nu$). Solid lines denote the reference friction curve for smooth pipe flow, equation~\eqref{eq:prandtl}. Symbols indicate present DNS data (circles), square duct DNS data of \citet{pinelli_10} (diamonds) and \citet{vinuesa_14} (squares), square duct experiments by \citet{jones_76} (stars) and \citet{leutheusser_63} (crosses), pipe flow DNS data of \citet{wu_08} (downtriangles), and pipe flow experiments by \citet{mckeon_04} (plus symbols). The figure inset shows a comparison with the predictions of the square-root area law~\citep{duan_12} (solid circles).}
  \label{fig:Cf}
 \end{center}
\end{figure}

The computed average friction factor, $f = 4 C_f = 8 {\tau}^*_w / (\rho u_b^2)$ is shown 
as a function of the bulk Reynolds number in figure~\ref{fig:Cf},
together with previous DNS data~\citep{pinelli_10,vinuesa_14}, and
assorted experimental data~\citep{jones_76}. Overall, the agreement with other DNS is quite good,
and results fall well within the scatter of the experimental data points.
It is a common notion in the engineering practice~\citep{schlichting_79}
that friction data for ducts with complex shape should collapse on
the friction curves for canonical flows, provided the bulk Reynolds 
number is constructed using a suitable length scale, the most frequently used
being the hydraulic diameter, defined as $D_h = 4 A/P$, 
with $A$ the cross-stream area and $P$ the duct perimeter. 
Hence, $D_h=2h$ for a square duct, and $\Rey_{D_h} = \Rey_b$. Several authors 
have noticed shortcomings of the hydraulic diameter concept, and proposed alternatives.
For instance, \citet{jones_76} analyzed data for rectangular ducts, and 
come to the conclusion that a corrective factor should be applied to the 
conventional hydraulic diameter, which happens to be $1.125$ for a square duct.
More recently, \citet{duan_12} proposed replacing the hydraulic diameter with the 
square-root of the cross- stream area. For the case of a square duct, this correction amounts
to multiplying the bulk Reynolds number by a factor $2/\sqrt{\pi}$, not far from
Jones' correction.
For the sake of checking the validity of the above semi-empirical 
formulations, in figure~\ref{fig:Cf} we alo report numerical and experimental 
data for pipe flow~\citep{mckeon_04,wu_08}, 
as well as the Karman-Prandtl theoretical friction law,
\begin{equation}
1/f^{1/2} = 2 \log_{10} (\Rey_{D_h} f^{1/2}) - 0.8. \label{eq:prandtl}
\end{equation}
Excellent agreement of the DNS data with equation~\eqref{eq:prandtl} 
is observed using the standard hydraulic diameter formulation, especially for 
the higher-$\Rey$ flow cases (C, D).
However, proposed corrections to the hydraulic diameter scaling also seem to work
very well (see the inset of figure~\ref{fig:Cf}), yielding perhaps better agreement at lower Reynolds number.
The success of the hydraulic diameter concept (as well as its variations) 
may be tentatively explained based on the empirical finding 
(recalling figure~\ref{fig:vel_prof_star}(b)) that 
the inner-scaled mean velocity profiles along much of the duct perimeter 
exhibit near invariance in the outer layer also when scaled with global wall units.
Hence, approximating the outer layer profiles with the classical log law,
namely $u^*=1/k \log y^* + C$, and integrating over an octant, 
the following expression for the bulk velocity results 
\begin{equation}
\frac {u_b}{u_{\tau}^*} = \frac{Q}{4 h^2 u_{\tau}^*} = 8 \int_0^h \int_0^z \left( \frac 1k \log y/h + \frac 1k \log \Rey_{\tau}^* + C  \right) \diff y \diff z = \frac 1k \log \Rey_{\tau}^* + C - \frac 3{2k} . \label{eq:ub_duct}
\end{equation}
The friction coefficient can then be evaluated as a function of $\Rey_b$ from~\eqref{eq:ub_duct} using
$C_f = 2 (u_{\tau}^*/u_b)^2$, $\Rey_{\tau}^* = \Rey_b u_{\tau}^* / (2 u_b)$.
Equation~\eqref{eq:ub_duct} should be compared with the corresponding expression for 
a circular duct
\begin{equation}
\frac {u_b}{u_{\tau}} = \frac 1k \log \Rey_{\tau} + C - \frac 3{2k}, \label{eq:ub_pipe}
\end{equation}
where $\Rey_{\tau} = D u_{\tau} / (2 \nu)$, and leading to equation~\eqref{eq:prandtl}.
The two expressions become identical provided $2 h = D$, hence provided 
the Reynolds numbers based on the hydraulic diameter are the same.
It is interesting that equation~\eqref{eq:ub_duct} is basically arrived at 
by neglecting the local wall shear stress variation along the duct perimeter,
and disregarding the flow deceleration at corners.
Apparently, these effects very nearly cancel out.

\section{A model for the secondary motions}

\begin{figure}
 \begin{center}
  (A)
  \includegraphics[width=4.0cm,clip]{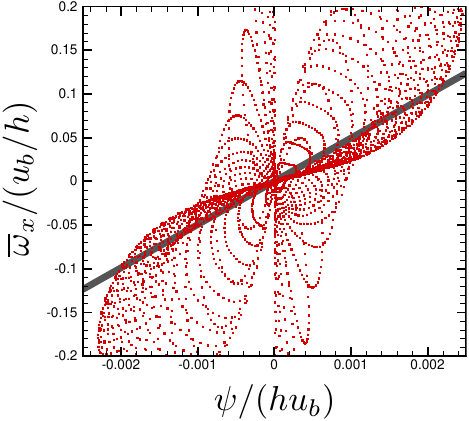}
  (B)
  \includegraphics[width=4.0cm,clip]{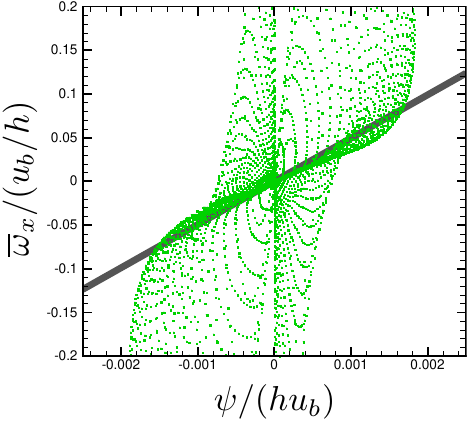} \\
  (C)
  \includegraphics[width=4.0cm,clip]{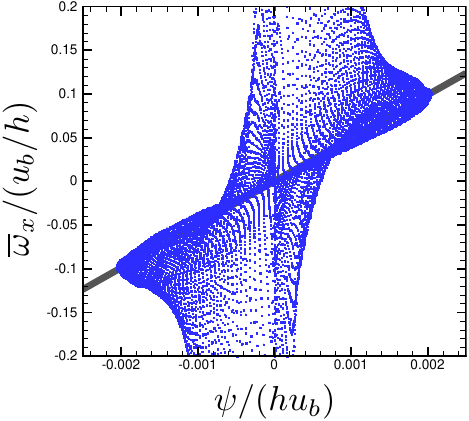}
  (D)
  \includegraphics[width=4.0cm,clip]{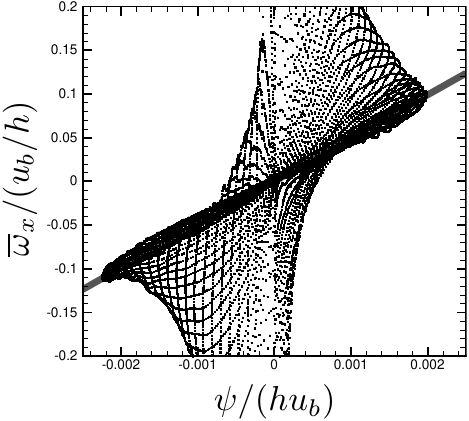}
  \caption{Scatter plots of cross-stream vorticity as a function of streamfunction.
The diagonal line denotes the trend $\overline{\omega}_x = k^2 \psi$, with $k^2=49.452$.}
  \label{fig:scatter}
 \end{center}
\end{figure}

\begin{table}
\centering
\begin{tabular}{ccccccccccc}
 Mode & 1 & 2 & 3 & 4 & 5 & 6 & 7 & 8 & 9 & 10 \\
\hline
$k^2$ & 4.936 & 12.343 & 19.756 & 24.700 & 32.120 & 42.021 & 44.497 & 49.452 & 61.848 & 64.326 \\
\hline
\end{tabular}
\caption{List of ten smallest eigenvalues of Laplace operator in a $[-1:1]\times[-1:1]$ square.}
\label{tab:eigs}
\end{table}%

\begin{figure}
 \begin{center}
  (a)
  \includegraphics[width=3.6cm,clip]{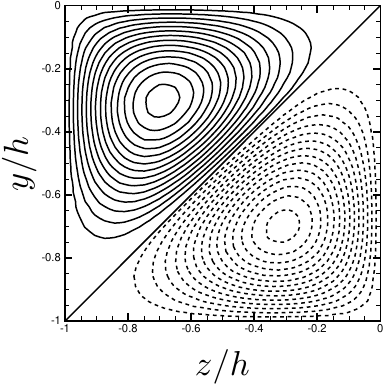}
  (b)
  \includegraphics[width=3.6cm,clip]{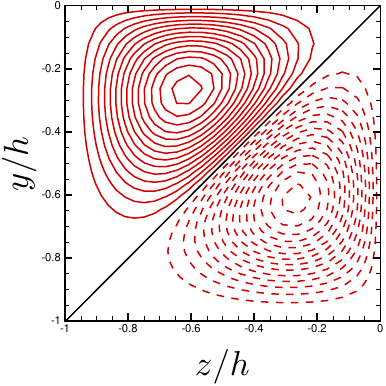}
  (c)
  \includegraphics[width=3.6cm,clip]{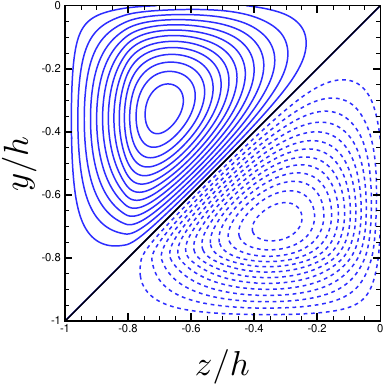}
  \caption{Eigenfunctions of equation~\eqref{eq:helmholtz} with $k^2=49.452$ (a),
  and of equation~\eqref{eq:biharm} with $\lambda^2=67.280$ (b). Panel (c) shows the streamlines for flow case D, upon removing the vorticity within a band of $200 \delta_v^*$ from the walls.}
  \label{fig:modes}
 \end{center}
\end{figure}

Based on the outcome of the previous sections, it is possible to formulate a tentative model
for the structure of the secondary motions. As shown in figure~\ref{fig:omx_budg}, the convective
terms in the streamwise vorticity budget are negligible except in the corner proximity.
Hence, the relevant condition for the existence of stationary cross-stream flow is
$\overline{\omega}_x = f(\psi)$~\citep{batchelor_69}. This ansatz can be empirically verified based on the DNS data. 
Scatter plots of cross-stream vorticity and streamfunction are shown in figure~\ref{fig:scatter},
for all points in the channel cross-section.
Despite the presence of significant dispersion, especially associated with near-wall points,
the data suggest a tendency for $\psi$ and $\overline{\omega}_x$ to become more and more correlated 
as $\Rey$ is increased, as already pointed out commenting figure~\ref{fig:psiomega}.
Specifically, a nonlinear relationship is initially identified, which then
yields a distinct linear variation at $\Rey_{\tau}^* \ge 500$.
For flow case D, we find that over $70\%$ of points in the duct cross-section is found within
a band of $\pm 0.03 u_b/h$ from the alleged linear relation.
Based on this evidence, we argue that secondary motions at sufficiently high $\Rey$ 
may be approximately characterized using equation~\eqref{eq:poisson} with the prescription $\overline{\omega}_x = k^2 \psi$,
which yields the classical Helmholtz equation for the cross-stream streamfunction,
\begin{equation}
(\nabla^2 + k^2) \psi = 0 , \label{eq:helmholtz}
\end{equation}
with $\psi=0$ at the duct boundaries.
Clearly, equation \eqref{eq:helmholtz} cannot also accommodate the no-slip condition
at the wall. However, we still expect that those solutions may be relevant in the 
high-Reynolds-number limit,
also in light of the previously noticed evidence that viscous effects 
tend to be progressively confined to the near-wall region and to the duct corners.
The cross-stream eddies would then correspond to eigenfunctions of equation~\eqref{eq:helmholtz},
and the admissible values of the constant $k^2$ would be the corresponding eigenvalues.
It is important that the steady equation for the streamfunction results from a time-evolving process which
is controlled by turbulence production and viscous diffusion. Hence, it may be expected that
in the long term, the only surviving mode(s) will be those which least damped 
from viscous diffusion, which occurs for those modes having smallest $k^2$. 
The first few eigenvalues of \eqref{eq:helmholtz} in a square domain 
are listed in table~\ref{tab:eigs}, some of them corresponding to multiple 
eigenfunctions. The DNS data (see figure~\ref{fig:scatter}) 
suggest that the mode which manifests itself is the one
corresponding to the eighth smallest eigenvalue. Notably, this corresponds to 
the first eigenmode to respect the eight-fold statistical symmetry properties for  square duct, and whose 
streamlines bisect the duct corners, hence respecting the physical requirement of 
being effective in transporting momentum from the bulk flow into
the corner region.
The Laplace eigenmode corresponding to $k^2=49.452$ 
is shown in figure~\ref{fig:modes}(a). For the sake of comparison, in panel (c) we also show the streamlines obtained for flow case D after removing 
the near-wall vorticity, which is consistent with the assumptions under 
which the theory has been developed.
Comparing figure~\ref{fig:modes}(c) with the unfiltered 
streamfunction given in figure~\ref{fig:psiomega}D shows that
the main effect of removing the near-wall vorticity is to have greater 
penetration of the streamlines into the corners.
The agreement of theory and filtered DNS field is apparently quite good.

It should be noted that a mathematical model for the secondary eddies in low-Reynolds number flow was 
also proposed by~\citet{wedin_08}, based on the concept of self-sustained process (SSP).
Based on that analysis, eigensolutions of the equation
\begin{equation}
(\nabla^4 + \lambda^2 \nabla^2)\psi=0 , \label{eq:biharm}
\end{equation}
with homogeneous boundary conditions for $\psi$ and its wall-normal derivative, 
should emerge as a result of selective viscous decay.
Specifically, eight eddies geometrically similar to those observed in developed
flow were recovered for one particular eigenmode (corresponding to $\lambda^2=67.280$) whose shape
is shown in figure~\ref{fig:modes}(b). 
Although this eigenmode is not qualitatively too different from the filtered DNS 
field of panel (c) (but note that the values of $\psi$ are here smaller near the walls as a consequence of the no-slip condition),
the $\psi$-$\omega_x$ signature (not reported) 
highlights clustering of data points around the $\overline{\omega}_x = \lambda^2 \psi$ line, which is quantitatively different than observed in DNS (recalling figure~\ref{fig:scatter}).

\begin{figure}
 \begin{center}
  (a)
  \includegraphics[height=3.6cm,clip]{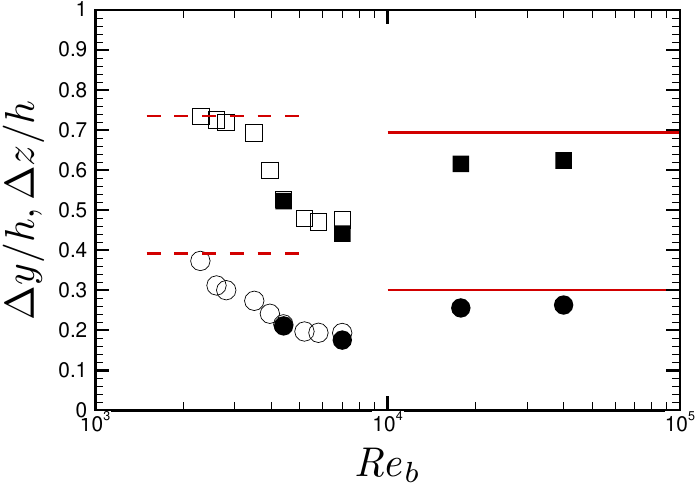}
  (b)
  \includegraphics[height=3.6cm,clip]{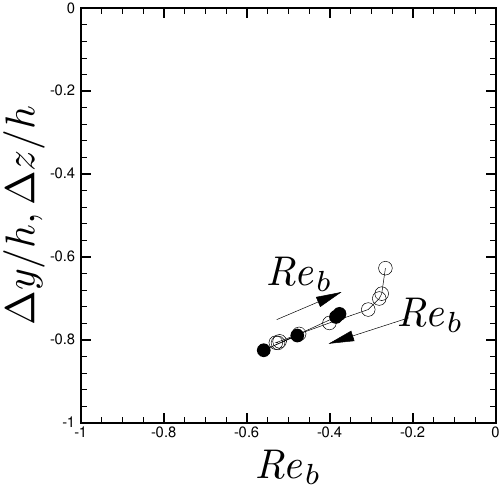}
  \caption{Position of vortex centers as a function of Reynolds number: 
  (a) distance from bottom wall (circles) and from left wall (squares); 
  (b) position in cross-stream plane.
  Solid symbols denote the present DNS data, open symbols denote DNS data of \citet{pinelli_10}.
  In panel (a) solid lines denote the vortex centers for eigensolutions of equation~\eqref{eq:helmholtz} with $k^2=49.452$, and dashed lines for eigensolutions of equation~\eqref{eq:biharm} with $k^2=67.280$.}
  \label{fig:centers}
 \end{center}
\end{figure}

More quantitative evaluation of the properties of the secondary eddies
can be gained from figure~\ref{fig:centers}, where we show the distance of the 
vortex centers from the duct walls as a function of the Reynolds number.
Low-Reynolds-number data from \citet{pinelli_10} are also shown.
A non-monotonic drift of the vortex center position is found at low Reynolds number,
whereby centers initially approach the corner, and then move away from it.
Interestingly, it appears that the predictions of the SSP theory~\citep{wedin_08} 
are in good agreement with DNS at low $\Rey$, whereas the Laplace eigenfunctions seem to 
yield the correct asymptotic trend at high Reynolds number.

\section{Conclusions} \label{sec:concl}

We have carried out a DNS study of turbulent flow in a square duct in an
unprecedentedly wide range of Reynolds numbers. The flow cases have been designed
and carried out in such a way as to minimize possible uncertainties associated with
limited grid resolution and/or lack of convergence of the statistical ensemble.
In our opinion, this allows to draw more solid conclusions regarding the 
structure and the dynamical effect of the secondary motions than possible in previous studies.
First, we find that the typical intensity of the secondary motions is very nearly unaffected
by Reynolds number variation, scaling with the bulk velocity (in fact, a small fraction of it).
Second, the secondary motions involve two coexisting circulations, with similar shape:
i) a corner circulation, whose typical length scale is the global viscous length
scale $\delta_v^*$, and ii) a core circulation, whose typical length scale is $h$.
Hence, the former is associated with higher values of the vorticity, but it becomes progressively 
confined to the duct corners as $\Rey$ is increased. The core circulation on the other
hand is weaker, but its effect is dominant over most of the duct cross-section.
The DNS data show that mean cross-stream convection plays a minor role in the streamwise 
vorticity budget, which points to a simple structure of the secondary motions
(at least away from walls), that involves a linear correlation between the
averaged vorticity and the streamfunction, again well documented from DNS data
at sufficiently high Reynolds number.
A simple analytical characterization of the cross-stream eddies also directly arises
under the assumption of negligible effect of mean convection,
as of particular eigenmodes of the Laplace operator. Comparison with the shape of 
eddies in DNS shows a tendency for the latter to assume features similar to those
theoretically predicted in the high-$\Rey$ limit, whereas they resemble those predicted
from viscous theories in the low-$\Rey$ end.
Of course, the intensity of the secondary flow is not predicted by the theory,
being the likely result of a time-evolutionary process involving the competing
effect of production from secondary stress gradients and viscous diffusion.

A quantitative characterization of the cross-stream eddies may be useful for certain purposes,
such as for instance validating RANS turbulence models, and perhaps to develop predictive 
models for diffusion of inertial particles. 
However, regarding their direct effect on the bulk flow properties (mainly, friction),
evidence reported in the present study seems to suggest that it is quite weak, if any.
In first instance, analysis of the mean streamwise momentum equation shows that,
with the exception of corners, cross-stream convection is much smaller than the turbulence
terms and viscous diffusion. Hence, near-equilibrium layers do form adjacent to each wall,
which exhibit very similar structure as the wall layers in canonical flows.
Cross-stream convection has the main effect of modulating the thickness
of those layers, hence yielding mild ($10\%$ at most) variation of the local wall friction along the 
duct perimeter. This effect probably further cancels out upon averaging over the 
duct perimeter, or (as is likely to be the case) because the flow away from walls 
responds to the imposed (spatially uniform) pressure gradient, rather than to the local
wall friction. Furthermore, the effect of 
velocity distortions in corners is very small when integrated over the cross-section.
As a result, it turns out that the mean velocity field can be characterized
with good accuracy in terms of universal profiles normal to each wall and
extending all the way to the duct corner bisector, with transition from 
wall scaling to pressure scaling occurring at a wall distance of $0.2h$.
The resulting predictive formula for the friction coefficient (equation~\eqref{eq:ub_duct})
is identical to the theoretical friction formula for pipes, thus providing
theoretical support for the concept of hydraulic diameter (and its variations) 
as a convenient way of incorporating the effect of the cross-stream geometry. 
For the case under scrutiny, refinements of the hydraulic diameter concept~\citep{jones_76,duan_12}
yield friction predictions with the same accuracy as the baseline parametrization.
Different geometries such as triangles, for which corrections are expected to be stronger, 
should be studied to identify the `best' predictive formula.

It would be of extreme interest to carry over the results of the present study
to yet higher Reynolds number, to further corroborate (or disproof) the findings.
In any case, we find that the study of flows in duct with complex shapes, other than 
being interesting for its own sake, may also help to shed light onto
phenomena of more canonical flows (e.g. pressure gradient or wall friction scaling),
which cannot be conventionally sorted out.


\begin{acknowledgments}
We wish to thank Dr. D. Biau for providing the eigenmodes of equation~\eqref{eq:biharm},
and Prof. A. Pinelli and Dr. R. Vinuesa for sharing their DNS data.
We further acknowledge that most of the results reported in this paper have been achieved using the PRACE Research Infrastructure resource MARCONI based at CINECA, Casalecchio di Reno, Italy.
\end{acknowledgments}
\bibliographystyle{jfm}
\bibliography{references}

\begin{thebibliography}{40}
\expandafter\ifx\csname natexlab\endcsname\relax\def\natexlab#1{#1}\fi

\bibitem[Adrian \& Marusic(2012)]{adrian_12}
{\sc Adrian, R.J. \& Marusic, I.} 2012 Coherent structures in flow over
  hydraulic engineering surfaces. {\em J.~Hydraulic~Res.\/} {\bf 50}, 451--464.

\bibitem[Batchelor(1969)]{batchelor_69}
{\sc Batchelor, G.} 1969 {\em An Introduction to Fluid Dynamics\/}. Cambridge
  University Press.

\bibitem[Bernardini {\em et~al.\/}(2014)Bernardini, Pirozzoli \&
  Orlandi]{bernardini_14}
{\sc Bernardini, M., Pirozzoli, S. \& Orlandi, P.} 2014 Velocity statistics in
  turbulent channel flow up to ${R}e_{\tau} = 4000$. {\em J.\ Fluid\ Mech.\/}
  {\bf 742}, 171--191.

\bibitem[Bradshaw(1987)]{bradshaw_87}
{\sc Bradshaw, P.} 1987 Turbulent secondary flows. {\em
  Annu.~Rev.~Fluid~Mech.\/} {\bf 19}, 53--74.

\bibitem[Brundrett \& Baines(1964)]{brundrett_64}
{\sc Brundrett, E. \& Baines, W.D.} 1964 The production and diffusion of
  vorticity in duct flow. {\em J. Fluid Mech.\/} {\bf 19}~(03), 375--394.

\bibitem[Demuren \& Rodi(1984)]{demuren_84}
{\sc Demuren, A.O. \& Rodi, W.} 1984 Calculation of turbulence-driven secondary
  motion in non-circular ducts. {\em J. Fluid Mech.\/} {\bf 140}, 189--222.

\bibitem[Duan {\em et~al.\/}(2012)Duan, Yovanovich \& Muzychka]{duan_12}
{\sc Duan, S., Yovanovich, M.M. \& Muzychka, Y.S.} 2012 Pressure drop for fully
  developed turbulent flow in circular and noncircular ducts. {\em
  J.~Fluids~Eng.\/} {\bf 134}~(6), 061201.

\bibitem[Einstein \& Li(1958)]{einstein_58}
{\sc Einstein, H.A. \& Li, H.} 1958 Secondary currents in straight channels.
  {\em Trans. Am. Geophys. Union\/} {\bf 39}~(6), 1085--1088.

\bibitem[Gavrilakis(1992)]{gavrilakis_92}
{\sc Gavrilakis, S.} 1992 Numerical simulation of low-{R}eynolds-number
  turbulent flow through a straight square duct. {\em J. Fluid Mech.\/} {\bf
  244}, 101--129.

\bibitem[Gessner \& Jones(1965)]{gessner_65}
{\sc Gessner, F.B. \& Jones, J.B.} 1965 On some aspects of fully-developed
  turbulent flow in rectangular channels. {\em J. Fluid Mech.\/} {\bf 23}~(04),
  689--713.

\bibitem[Hoagland(1960)]{hoagland_60}
{\sc Hoagland, L.C.} 1960 Fully developed turbulent flow in straight
  rectangular ducts -- secondary flow, its cause and effect on the primary
  flow. PhD thesis, Department of Mechanical Engineering, Massachusetts
  Institute of Technology.

\bibitem[Huser \& Biringen(1993)]{huser_93}
{\sc Huser, A. \& Biringen, S.} 1993 Direct numerical simulation of turbulent
  flow in a square duct. {\em J. Fluid Mech.\/} {\bf 257}, 65--95.

\bibitem[Jones(1976)]{jones_76}
{\sc Jones, O.C.} 1976 An improvement in the calculation of turbulent friction
  in rectangular ducts. {\em ASME~J.~Fluids~Engng.\/} {\bf 98}, 173--181.

\bibitem[Kim \& Moin(1985)]{kim_85}
{\sc Kim, J. \& Moin, P.} 1985 Application of a fractional-step method to
  incompressible {Navier-Stokes} equations. {\em J. Comput. Phys.\/} {\bf 59},
  308--323.

\bibitem[Launder \& Ying(1972)]{launder_72}
{\sc Launder, B.E \& Ying, W.M.} 1972 Secondary flows in ducts of square
  cross-section. {\em J. Fluid Mech.\/} {\bf 54}~(02), 289--295.

\bibitem[Leutheusser(1963)]{leutheusser_63}
{\sc Leutheusser, H.J.} 1963 Turbulent flow in rectangular ducts. {\em
  J.~Hydr.~Div.~ASCE\/} {\bf 89}~(3), 1--19.

\bibitem[Leutheusser(1984)]{leutheusser_84}
{\sc Leutheusser, H.~J.} 1984 Velocity distribution and skin friction
  resistance in rectangular ducts. {\em J.~Wind~Eng.~Ind.~Aero.\/} {\bf 16},
  315--327.

\bibitem[Mani {\em et~al.\/}(2013)Mani, Babcock, Winkler \& Spalart]{mani_13}
{\sc Mani, M., Babcock, D., Winkler, C. \& Spalart, P.} 2013 Predictions of a
  supersonic turbulent flow in a square duct. In {\em 51st AIAA Aerospace
  Sciences Meeting\/}, p. 860.

\bibitem[Marin {\em et~al.\/}(2016)Marin, Vinuesa, Obabko \&
  Schlatter]{marin_16}
{\sc Marin, O., Vinuesa, R., Obabko, A.V. \& Schlatter, P.} 2016
  Characterization of the secondary flow in hexagonal ducts. {\em Phys.
  Fluids\/} {\bf 28}~(12), 125101.

\bibitem[Mckeon {\em et~al.\/}(2004)Mckeon, Li, Jiang, Morrison \&
  Smits]{mckeon_04}
{\sc Mckeon, B.J., Li, J., Jiang, W., Morrison, J.F. \& Smits, A.J.} 2004
  Further observations on the mean velocity distribution in fully developed
  pipe flow. {\em J.~Fluid~Mech.\/} {\bf 501}, 135--147.

\bibitem[Modesti \& Pirozzoli(2016{\natexlab{{\em a\/}}})]{modesti_16a}
{\sc Modesti, D. \& Pirozzoli, S.} 2016{\natexlab{{\em a\/}}} An efficient
  semi-implicit solver for direct numerical simulation of compressible flows at
  all speeds. {\em arXiv preprint arXiv:1608.08513\/} .

\bibitem[Modesti \& Pirozzoli(2016{\natexlab{{\em b\/}}})]{modesti_16}
{\sc Modesti, D. \& Pirozzoli, S.} 2016{\natexlab{{\em b\/}}} Reynolds and
  {M}ach number effects in compressible turbulent channel flow. {\em Int. J.
  Heat Fluid Flow\/} {\bf 59}, 33--49.

\bibitem[Nezu(2005)]{nezu_05}
{\sc Nezu, I.} 2005 Open-channel flow turbulence and its research prospect in
  the 21st century. {\em J.~Hydraulic~Eng.\/} {\bf 131}, 229--246.

\bibitem[Nikuradse(1930)]{nikuradse_30}
{\sc Nikuradse, J.} 1930 Turbulente str\"omung in nicht-kreisf\"ormigen rohren.
  {\em Ing.~Arch.\/} {\bf 1}, 306--332.

\bibitem[Oliver {\em et~al.\/}(2014)Oliver, Malaya, Ulerich \&
  Moser]{oliver_14}
{\sc Oliver, T.A., Malaya, N., Ulerich, R. \& Moser, R.D.} 2014 Estimating
  uncertainties in statistics computed from direct numerical simulation. {\em
  Phys. Fluids\/} {\bf 26}~(3), 035101.

\bibitem[Orlandi(1990)]{orlandi_90}
{\sc Orlandi, P.} 1990 Vortex dipole rebound from a wall. {\em Phys. Fluids\/}
  {\bf 2}, 1429--1436.

\bibitem[Orlandi(2012)]{orlandi_12}
{\sc Orlandi, P.} 2012 {\em Fluid flow phenomena: a numerical toolkit\/}, ,
  vol.~55. Springer Science \& Business Media.

\bibitem[Pinelli {\em et~al.\/}(2010)Pinelli, Uhlmann, Sekimoto \&
  Kawahara]{pinelli_10}
{\sc Pinelli, A., Uhlmann, M., Sekimoto, A. \& Kawahara, G.} 2010 Reynolds
  number dependence of mean flow structure in square duct turbulence. {\em J.
  Fluid Mech.\/} {\bf 644}, 107--122.

\bibitem[Pirozzoli(2010)]{pirozzoli_10}
{\sc Pirozzoli, S.} 2010 Generalized conservative approximations of split
  convective derivative operators. {\em J. Comput. Phys.\/} {\bf 229}~(19),
  7180--7190.

\bibitem[Pirozzoli \& Bernardini(2013)]{pirozzoli_13}
{\sc Pirozzoli, S. \& Bernardini, M.} 2013 Probing high-{Reynolds}-number
  effects in numerical boundary layers. {\em Phys. Fluids\/} {\bf 25}, 021704.

\bibitem[Pope(2000)]{pope_00}
{\sc Pope, S.B.} 2000 {\em Turbulent flows\/}. Cambridge University Press.

\bibitem[Prandtl(1927)]{prandtl_27}
{\sc Prandtl, L.} 1927 {\"U}ber die ausgebildete turbulenz. In {\em
  Verh.~2nd~Int.~Kong.~f\"ur Tech.~Mech.\/}.

\bibitem[Schlichting(1979)]{schlichting_79}
{\sc Schlichting, H.} 1979 Boundary layer theory. {\em McGraw-Hill, New York\/}
  .

\bibitem[Speziale(1982)]{speziale_82}
{\sc Speziale, C.G.} 1982 On turbulent secondary flows in pipes of noncircular
  cross-section. {\em Int. J. Eng. Sci.\/} {\bf 20}~(7), 863--872.

\bibitem[Uhlmann {\em et~al.\/}(2007)Uhlmann, Pinelli, Kawahara \&
  Sekimoto]{uhlmann_07}
{\sc Uhlmann, M., Pinelli, A., Kawahara, G. \& Sekimoto, A.} 2007 Marginally
  turbulent flow in a square duct. {\em J. Fluid Mech.\/} {\bf 588}, 153--162.

\bibitem[Vinuesa {\em et~al.\/}(2014)Vinuesa, Noorani, Lozano-Dur{\'a}n,
  Khoury, Schlatter, Fischer \& Nagib]{vinuesa_14}
{\sc Vinuesa, R., Noorani, A., Lozano-Dur{\'a}n, A., Khoury, G.K.E., Schlatter,
  P., Fischer, P.F. \& Nagib, H.M.} 2014 Aspect ratio effects in turbulent duct
  flows studied through direct numerical simulation. {\em J. Turbulence\/} {\bf
  15}~(10), 677--706.

\bibitem[Vinuesa {\em et~al.\/}(2016)Vinuesa, Prus, Schlatter \&
  Nagib]{vinuesa_16}
{\sc Vinuesa, R., Prus, C., Schlatter, P. \& Nagib, H.M.} 2016 Convergence of
  numerical simulations of turbulent wall-bounded flows and mean cross-flow
  structure of rectangular ducts. {\em Meccanica\/} {\bf 51}~(12), 3025--3042.

\bibitem[Wedin {\em et~al.\/}(2008)Wedin, Biau, Bottaro \& Nagata]{wedin_08}
{\sc Wedin, H., Biau, D., Bottaro, A. \& Nagata, M.} 2008 Coherent flow states
  in a square duct. {\em Phys. Fluids\/} {\bf 20}~(9), 094105.

\bibitem[Wu \& Moin(2008)]{wu_08}
{\sc Wu, X. \& Moin, P.} 2008 A direct numerical simulation study on the mean
  velocity characteristics in turbulent pipe flow. {\em J. Fluid Mech.\/} {\bf
  608}, 81--112.

\bibitem[Zhang {\em et~al.\/}(2015)Zhang, Trias, Gorobets, Tan \&
  Oliva]{zhang_15}
{\sc Zhang, H., Trias, F.X., Gorobets, A., Tan, Y. \& Oliva, A.} 2015 Direct
  numerical simulation of a fully developed turbulent square duct flow up to
  ${R}e_{\tau}= 1200$. {\em Int. J. Heat Fluid Flow\/} {\bf 54}, 258--267.

\end{thebibliography}
\end{document}